\documentclass[fp,twocolumn]{jpsj3}
\usepackage{graphicx, color}
\usepackage{physics}
\usepackage{bm}
\usepackage{mathtools}
\usepackage{newtx}
\usepackage[version=3]{mhchem}
\usepackage[normalem]{ulem}
\usepackage{braket}
\usepackage{comment}
\usepackage{anyfontsize}
\usepackage{cite}
\usepackage{cuted}

\newcommand{\so}{\Delta_{\mathrm{so}}}
\newcommand{\AM}{\Delta_{\mathrm{AM}}}
\newcommand{\kF}{k_{\mathrm{F}}}
\newcommand{\Tc}{T_{\mathrm{c}}}

\DeclareRobustCommand{\Erase}{\bgroup\markoverwith{\textcolor{red}{\rule[0.5ex]{2pt}{1.0pt}}}\ULon}

\title{Finite-momentum superconductivity in two-dimensional altermagnets with a Rashba-type spin-orbit coupling}

\newcommand{\TohokuUniv}{Department of Applied Physics, Tohoku University, Sendai 980-8579, Japan}

\author{Kohei Mukasa$^1$\thanks{mukasa.kohei.p7@dc.tohoku.ac.jp}, and Yusuke Masaki$^{1,2}$\thanks{yusuke.masaki.c1@tohoku.ac.jp}}
\inst{$^1$\TohokuUniv, \\
$^2$Research and Education Center for Natural Sciences, Keio University, Hiyoshi 4-1-1, Yokohama, Kanagawa 223-8521, Japan}

\date{\today}

\abst{
We theoretically study finite-momentum superconductivity in two-dimensional (2D) altermagnets with a Rashba-type spin-orbit coupling (RSOC).
We show the phase diagrams obtained by solving a linearized gap equation, considering two directions of the N\'{e}el vector of the 2D altermagnet: parallel to the $xy$ plane (in-plane) and perpendicular to the $xy$ plane (out-of-plane).
For the in-plane N\'{e}el vector, we find two different finite-momentum $d_{x^2-y^2}$-wave superconducting states distinguished by a dominant pairing channel: an inter-band pairing or an intra-band pairing.
Furthermore, it is shown that an anisotropic deformation of Fermi surfaces caused by spin-splitting effects due to the in-plane N\'{e}el vector and the RSOC can contribute to the stabilization of the finite-momentum superconductivity.
We also perform the self-consistent calculations, and verify the existence of these superconducting states in addition to identifying other possible superconducting states and transition lines away from the phase boundary.
For the out-of-plane N\'{e}el vector, the finite-momentum superconductivity is realized only in the inter-band pairing mechanism, which is in contrast to the in-plane case.
}

\begin{document}
\maketitle
\section{Introduction}
The formation of Cooper pairs is a fundamental concept to understand distinctive properties of superconductors.
In the conventional superconductors, the Cooper pair is formed by two electrons of the time reversal pair on the Fermi surface~(FS), and thus its center-of-mass momentum becomes zero.
However, once a spin degeneracy of the energy dispersion is lifted by the breaking of the time reversal symmetry, finite-momentum superconductivity, the Cooper pairs with a finite momentum $\bm{q}\neq \bm{0}$, can be stabilized.
The finite-momentum superconductivity induced by a strong magnetic field or an intrinsic net magnetization is known as the Fulde--Ferrell--Larkin--Ovchinnikov (FFLO) state\cite{FFstate,LOstate,FFLO-review-by-Shimahara}.
The LO state is characterized by two nonzero momenta $\pm \bm{q}$ and exhibits an amplitude modulation of the order parameter, while the FF state chooses a single nonzero momentum $\bm{q}$ and exhibits a phase modulation of the order parameter.

In noncentrosymmetric systems, an antisymmetric spin-orbit coupling such as a Rashba-type spin-orbit coupling (RSOC) can arise. There are a lot of studies on superconductors with a spin-orbit coupling~\cite{Barzykin_Gor’kov_2002,Kaur_Agterberg_Sigrist_2005,Agterberg-helicalstripe-2007, Dimitrova_Feigel’man_2007,Yanase_Sigrist_2008, Samokhin_2008,Michaeli_Potter_Lee_2012, helical_Zwicknagl_Jahns_Fulde_2017, topological-metal-liang-fu-2021,Aoyama_stripe-order, Smidman_Salamon_Yuan_Agterberg_2017}. 
Especially, coupled with an in-plane magnetic field, the RSOC can make the FF state more stable than the LO state because the two momenta $\pm \bm{q}$ is no longer equivalent. In this context, the FF state is called the  helical state~\cite{Kaur_Agterberg_Sigrist_2005, Agterberg-helicalstripe-2007, helical_Zwicknagl_Jahns_Fulde_2017, topological-metal-liang-fu-2021}.
Because both of the spatial inversion symmetry and the time reversal symmetry are broken, nonreciprocal transports such as the superconducting diode effect (SDE) can be realized~\cite{Nagaosa_Yanase_2024,SDE-Yuan_Fu_2022,phennomelogical-SDE-He_Tanaka_Nagaosa_2022,daido_diode,PhysRevB-daido-diode,Banerjee_Scheurer_2024,Hasan_Shaffer_Khodas_Levchenko_2024}.
Since the helical superconducting state is believed to be a key to the SDE \cite{daido_diode,phennomelogical-SDE-He_Tanaka_Nagaosa_2022,SDE-Yuan_Fu_2022,Nakamura_Daido_Yanase_2024, Ilić_Bergeret_2022}, a fundamental understanding of the finite-momentum superconductivity becomes more important to comprehend and to control nonreciprocal transport properties in superconductors.

Here, we focus on altermagnets as candidate materials for realization of the finite-momentum superconductivity. 
Altermagnetism is the newly discorvered magnetism which have a momentum-dependent spin-splitting of the energy band without net magnetization\cite{RuO2FS_Ahn_Hariki_Lee_Kuneš_2019, anisotropic_spin_Hayami_Yanagi_Kusunose_2019,AHE_Šmejkal_González-Hernández_Jungwirth_Sinova_2020,Yuan_Wang_Luo_Rashba_Zunger_2020,AHE_Naka_Hayami_Kusunose_Yanagi_Motome_Seo_2020, Hayami_Yanagi_Kusunose_2020, Šmejkal_Sinova_Jungwirth_2022, Šmejkal_Sinova_Jungwirth_2022_Emerging,McClarty_Rau_2024,spincurrent_generation_Naka_Hayami_Kusunose_Yanagi_Motome_Seo_2019}.
As the influence of magnetism on superconductivity has been broadly studied in condensed matter physics~\cite{Scalapino_2012,Fradkin_Kivelson_Tranquada_2015, Aoki_Ishida_Flouquet_2019}, the interplay between the altermagnetic spin-splitting and the superconductivity has recently attractted much attention\cite{SCinAM_topoSC,SCinAM_minimal-model,SCinAM_magnetoelectric,zero-field_Chakraborty_Black-Schaffer_2024,SCinAM_kappa-type,SCinAM_proximitized_effect,FMSC_in_AM_Sim_Knolle_2024, BFS_in_AM_Hong_Park_Kim_2024,SDE_in_AM_Banerjee,Bose_Vadnais_Paramekanti_2024,Giil_Linder_2024,TopoSC_Lee_Qian_Yang_2024, topoSC_Li_2024, Soto-Garrido_Fradkin_2014, Li_Liu_2023_topoSCinAM, Ghorashi_Hughes_Cano_2024, Chakraborty_Black-Schaffer_2024_constraint, Maeda_Fukaya_Yada_Lu_Tanaka_Cayao_2025, Chatterjee_Juričić_2025}.
Especially, the finite-momentum superconductivity in altermagnets has been theoretically proposed and shows some intriguing features such as the field-induced superconductivity\cite{zero-field_Chakraborty_Black-Schaffer_2024} and the zero-field FFLO state\cite{zero-field_Chakraborty_Black-Schaffer_2024,SCinAM_kappa-type,FMSC_in_AM_Sim_Knolle_2024}.
In addition, the SDE in altermagnets has been  reported in several theoretical works\cite{FMSC_in_AM_Sim_Knolle_2024, PerfectSDE_Chakraborty_Black-Schaffer_2024, SDE_in_AM_Banerjee}.
However, understanding of the property of the finite-momentum superconductivity in the altermagnet with the RSOC has been limited~\cite{SDE_in_AM_Banerjee}. 
It is necessary to understand the combined effect of the altermagnet and the RSOC when considering the application of altermagnets to the devices related to the SDE.

Motivated by these backgrounds, in this paper, we calculate the possibility of the $d$-wave finite-momentum superconductivity in a two-dimensional (2D) altermagnet mainly within a linearized gap equation, taking the effect of the RSOC into account.
Two directions of the N\'{e}el vector, in-plane and out-of-plane, are considered.
For the in-plane altermagnet, since the altermagnetic spin-splitting and the RSOC are coupled, the FSs are deformed into anisotropic shapes, resulting in two different finite-momentum superconducting states characterized by a dominant pairing channel: an inter-band pairing or an intra-band pairing. 
We explain stabilization mechanisms of the finite-momentum superconductivity on the basis of the shapes of the FSs. 
Since the validity of the linearized gap equation is restricted to regions near the phase boundary between superconducting and normal states, we also perform self-consistent calculations to investigate other possible superconducting states and transition lines farther from this boundary.

For the out-of-plane altermagnet, we find that the Bardeen--Cooper--Schrieffer (BCS) state with zero momentum is stable for the intra-band pairing and the finite-momentum superconductivity appears only for the inter-band pairing. The difference  between the directions of the N\'{e}el vector originates from qualitative changes of the FSs and spin textures.

\section{Model and Method}
\label{sec:model and method}
We consider finite-momentum superconductivity in metallic $d$-wave altermagnets with the RSOC. 
The normal-state Hamiltonian of this system is written as
\begin{equation} \label{eq:Total_Hamiltonian}
    \begin{aligned}
        \hat{H}_0 = &\sum_{\bm{k}}\sum_{\sigma\sigma'}\hat{c}^{\dagger}_{\bm{k}\sigma}[H_0(\bm{k})]_{\sigma\sigma'}\hat{c}_{\bm{k}\sigma'},
    \end{aligned}
\end{equation}
where $c^{\dagger}_{\bm{k}\sigma}$ ($c_{\bm{k}\sigma}$) is the creation (annihilation) operator of the electron with spin $\sigma$ and momentum $\bm{k}$. 
The $2 \times 2$ matrix $H_0(\bm{k})$ is given by
\begin{equation} \label{eq:H0}
    H_0(\bm{k}) = \frac{k^2}{2m} - \mu + \bm{h}(\bm{k})\cdot\bm{\sigma}, 
\end{equation}
where $m$ is the mass of the electron, $\mu$ is the chemical potential, and $\bm{\sigma}=(\sigma_x, \sigma_y, \sigma_z)$ is the Pauli vector. For later convenience, we introduce $\xi_k = k^2/(2m) - \mu$. In the last term, $\bm{h}(\bm{k})$ represents the momentum-dependent spin-splitting caused by the RSOC and the $d$-wave altermagnet, written as
\begin{equation} \label{eq:hvector}
    \bm{h}(\bm{k}) = \so\hat{\bm{k}}\times\bm{e}_z + \AM(\hat{k}_x^2-\hat{k}_y^2)\bm{n},
\end{equation}
where $\hat{\bm{k}} = \bm{k}/\kF$ with $\kF=\sqrt{2m\mu}$, $\so$ and $\AM$ are the strength of the RSOC and the $d$-wave altermagnetic spin-splitting, respectively. 
The N\'{e}el vector of the altermagnet is denoted by $\bm{n}$, and in this paper, we consider both cases: $\bm{n}$ is parallel to the $x$$y$ plane (the in-plane altermagnet), and $\bm{n}$ is along the $z$-axis (the out-of-plane altermagnet). 
The normal-state Hamiltonian \eqref{eq:Total_Hamiltonian} is transformed into the band representation $\hat{H}_0 = \sum_{\bm{k}}\sum_{\lambda=\pm}E_{\lambda}(\bm{k})\hat{b}_{\bm{k}\lambda}^{\dagger}\hat{b}_{\bm{k}\lambda}$, 
where $E_{\lambda}(\bm{k})$ is the eigenvalue of Eq.~\eqref{eq:H0} with the band index $\lambda(=\pm)$ and can be written as
\begin{equation} \label{eq:dispersion}
    E_{\lambda}(\bm{k}) = \xi_k + \lambda|\bm{h}(\bm{k})|.
\end{equation}
The corresponding eigenvector $\bm{w}_{\lambda}(\bm{k})$ gives the relation $\hat{c}_{\bm{k}\sigma}= \sum_{\lambda}[\bm{w}_{\lambda}(\bm{k})]_{\sigma}\hat{b}_{\bm{k}\lambda}$. These notations are used also in Sec.~\ref{sec:self-consistent_AM}.

Next, we introduce the following form of an attractive interaction for Cooper pair formations:
\begin{align}
    \hat{H}_V = \frac{1}{\Omega}\sum_{\bm{k}\bm{k}'}\hat{c}^{\dagger}_{\bm{k}+\bm{q}/2\uparrow}\hat{c}^{\dagger}_{-\bm{k}+\bm{q}/2\downarrow}V(|\bm{k}-\bm{k}'|)\hat{c}_{-\bm{k}'+\bm{q}/2\downarrow}\hat{c}_{\bm{k}'+\bm{q}/2\uparrow}, \label{eq:attractive_interaction}
\end{align}
where $\bm{q}$ is the center-of-mass momentum of the Cooper pairs, $\Omega$ is the volume of the system and $V(|\bm{k}-\bm{k}'|)$ is the Fourier component of the attractive interaction between two electrons.

We apply the mean-field approximation to Eq.~\eqref{eq:attractive_interaction} by defining the superconducting order parameter defined as
\begin{equation}\label{eq:def_Delta}
    \Delta_{\sigma\sigma'}(\bm{k}, \bm{q})=-\frac{1}{\Omega}\sum_{\bm{k}'}V(|\bm{k}-\bm{k}'|)\ev{\hat{c}_{\bm{k}'+\bm{q}/2\sigma}\hat{c}_{-\bm{k}'+\bm{q}/2\sigma'}}.
\end{equation}
Then, the total Hamiltonian $\hat{H}_0 + \hat{H}_V$ is reduced to
\begin{equation} \label{eq:MF_Hamiltonian}
    \begin{aligned}
        \hat{H}^{\mathrm{MF}} &= \sum_{\bm{k}}\sum_{\sigma\sigma'}\hat{c}^{\dagger}_{\bm{k}\sigma}[H_0(\bm{k})]_{\sigma\sigma'}\hat{c}_{\bm{k}\sigma'} \\
        &+\sum_{\bm{k}}\left(
        \Delta_{\uparrow\downarrow}(\bm{k},\bm{q})\hat{c}^{\dagger}_{\bm{k}+\bm{q}/2\uparrow}\hat{c}^{\dagger}_{-\bm{k}+\bm{q}/2\downarrow} + \mathrm{H.c.} \right)+ \mathrm{const.}
    \end{aligned}
\end{equation}
We linearize the gap equation \eqref{eq:def_Delta} 
by using the Gor'kov equation.
The Green's function and the anomalous Green's function are defined as follows:
\begin{align}
    \label{eq:def_G}
    G_{\sigma\sigma'}(\bm{k}_+,\omega_n) &\equiv -\int^{\beta}_0\mathrm{d}\tau\,e^{i\omega_n\tau}\ev{\hat{c}_{\bm{k}_+\sigma}(\tau)\hat{c}^{\dagger}_{\bm{k}_+\sigma'}}, \\
    \label{eq:def_F}
    F_{\sigma\sigma'}(\bm{k},\bm{q},\omega_n) &\equiv -\int^{\beta}_0\mathrm{d}\tau\,e^{i\omega_n\tau}\ev{\hat{c}_{\bm{k}_+\sigma}(\tau)\hat{c}_{-\bm{k}_-\sigma'}},
\end{align}
where $\bm{k}_{\pm}=\bm{k}\pm\bm{q}/2$, $\omega_n=\pi k_{\mathrm{B}}T(2n+1)$ are the Matsubara frequencies, $\beta=1/k_{\mathrm{B}}T$ is the inverse temperature, and $\hat{c}_{\bm{k}\sigma}(\tau)=e^{\tau\hat{H}^{\mathrm{MF}}}\hat{c}_{\bm{k}\sigma}e^{-\tau\hat{H}^{\mathrm{MF}}}$. The Gor'kov equation can be derived as
\begin{align} 
        [i\omega_n \check{I} - \check{H}_{\mathrm{BdG}}(\bm{k},\bm{q})]
        \check{G}(\bm{k},\bm{q};\omega_n)
        =
        \check{I} \label{eq:gor'kov eq}
\end{align}
with
\begin{align}
        \check{H}_{\mathrm{BdG}}(\bm{k},\bm{q})&=
        \begin{pmatrix}
            H_0(\bm{k}_+) & \Delta(\bm{k},\bm{q}) \\ 
            -\Delta^{*}(-\bm{k},\bm{q}) & -H^{*}_0(-\bm{k}_{-})
        \end{pmatrix},\label{eq:BdG hamiltonian}\\
        \check{G}(\bm{k},\bm{q};i\omega_n) &= 
        \begin{pmatrix}
           G(\bm{k}_+, \omega_n)  & F(\bm{k},\bm{q}, \omega_n) \\
            -F^{*}(-\bm{k},\bm{q}, \omega_n) & -G^{*}(-\bm{k}_-, \omega_n)
        \end{pmatrix},\label{eq:green's function matrix}
\end{align}
where $\check{\cdot}$ describes a $4\times 4$ matrix, and $\check{I}$ denotes the $4 \times 4$ unit matrix.
The matrix elements of $2\times 2$ matrices $\Delta(\bm{k},\bm{q})$, $G(\bm{k}_+,\omega_n)$, and $F(\bm{k},\bm{q},\omega_n)$ are given by Eq.~\eqref{eq:def_Delta}, 
Eq.~\eqref{eq:def_G}, and Eq.~\eqref{eq:def_F}, respectively.
Using the anomalous Green's function, the gap equation is represented as 
\begin{equation} \label{eq:Delta_&_F}
    \Delta(\bm{k},\bm{q}) = \frac{k_{\mathrm{B}}T}{\Omega}
    \sum_{\bm{k}'} \sum_{n} V(|\bm{k}-\bm{k}'|)F(\bm{k}', \bm{q},\omega_n).
\end{equation}

It has been pointed that $s$-wave symmetry does not favor the finite-momentum superconductivity in the $d$-wave altermagnet\cite{zero-field_Chakraborty_Black-Schaffer_2024,Soto-Garrido_Fradkin_2014, Chakraborty_Black-Schaffer_2024_constraint}. Since we are interested in the finite-momentum superconductivity, we consider the spin-singlet $d$-wave superconducting order parameter.
In Appendix~\ref{sec:uniform magnetic field}, we also consider the spin-singlet $s$-wave superconductivity in the presence of a uniform magnetic field instead of the altermagnetism for comparison.
The spin-singlet order parameter $\Delta(\bm{k}, \bm{q})$ and the attractive interaction $V(|\bm{k}-\bm{k}'|)$ can be decomposed as
\begin{align}
    \Delta(\bm{k}, \bm{q}) &= \Delta(\bm{q}) \psi_{\bm{k}}\, (i\sigma_y), \label{eq:Delta}\\ V(|\bm{k}-\bm{k}'|) &= -V\psi_{\bm{k}}\psi_{\bm{k}'}, \label{eq:Vk}
\end{align}
where $V>0$ is a constant attractive strength. Here, $\psi_{\bm{k}}=\sqrt{2}\cos{2\phi}$ is the form factor for the $d$-wave superconducting order parameter, where $\phi$ is an azimuthal angle in the $k_x$-$k_y$ plane. 
The node directions of the $\psi_{\bm{k}}$ is the same as those for the altermagnetic term in Eq.~\eqref{eq:hvector}.
From Eq.~\eqref{eq:gor'kov eq}, $F(\bm{k},\bm{q},i\omega_n)$ can be obtained perturbatively up to the first order of $\Delta(\bm{k}, \bm{q})$. 
Substituting the perturbative solution of $F(\bm{k},\bm{q},i\omega_n)$, Eqs.~\eqref{eq:Delta} and \eqref{eq:Vk} into Eq.~\eqref{eq:Delta_&_F},  we obtain the linearized gap equation
\begin{equation} \label{eq:gap_eq_dwave}
    1 = \frac{k_{\text{B}}TV}{2\Omega}
    \sum_{\bm{k}} \sum_{n} \Tr \sigma_y G_0(\bm{k}_+, \omega_n)\sigma_y G^{*}_0(-\bm{k}_-, \omega_n) \psi_{\bm{k}}^2,
\end{equation}
where $G_0(\bm{k})=(i\omega_n-H_0(\bm{k}))^{-1}$. By performing the summation over the Matsubara frequencies, Eq.~\eqref{eq:gap_eq_dwave} can be rewritten as~\cite{helical_Zwicknagl_Jahns_Fulde_2017,topological-metal-liang-fu-2021}
\begin{equation} \label{eq:def_Kernel}
    1 = \frac{V}{\Omega} \sum_{\bm{k}} \sum_{\lambda,p}
    \frac{|u_p|^2}{4E^{(s)}_{\lambda,p}}
    \frac{\psi_{\bm{k}}^2\sinh{\beta E^{(s)}_{\lambda,p}}}{\cosh{\beta E^{(s)}_{\lambda,p}}+\cosh{\beta E^{(a)}_{\lambda,p}}},
\end{equation}
where $\lambda=\pm$, and $p=\mathrm{inter}$ or $\mathrm{intra}$. Here, we have defined $E^{(s)}_{\lambda,p}$ and $E^{(a)}_{\lambda,p}$ as
\begin{align}
    E^{(s)/(a)}_{\lambda,\mathrm{inter}} &= \frac{E_{\lambda}(\bm{k}_+)\pm E_{-\lambda}(-\bm{k}_-)}{2} \\
    E^{(s)/(a)}_{\lambda,\mathrm{intra}} &= \frac{E_{\lambda}(\bm{k}_+)\pm E_{\lambda}(-\bm{k}_-)}{2},
\end{align}
and $|u_{\mathrm{inter/intra}}|^2$ as
\begin{equation}
    |u_{\mathrm{inter/intra}}|^2 = \frac{1\pm\cos{\theta}(\bm{k},\bm{q})}{2},
\end{equation}
where $\theta(\bm{k}, \bm{q})$ is the angle between $\bm{h}(\bm{k}_+)$ and $\bm{h}(-\bm{k}_-)$, that is,
\begin{equation}
    \cos{\theta}(\bm{k},\bm{q}) = \frac{\bm{h}(\bm{k}_+)\cdot\bm{h}(-\bm{k}_-)}{|\bm{h}(\bm{k}_+)||\bm{h}(-\bm{k}_-)|}.
\end{equation}

Note that
$|u_{\mathrm{inter}}|^2 = 1$ and $|u_{\mathrm{intra}}|^2 = 0$
($|u_{\mathrm{inter}}|^2 = 0$ and $|u_{\mathrm{intra}}|^2 = 1$)
for $\theta = 0$ ($\theta = \pi$).These relations indicate that the electron spins at $\bm{k}_+$ and $-\bm{k}_-$ are antiparallel (parallel) on the two different bands (the same band) for $\theta = 0$, because $\bm{h}(\bm{k})$ is parallel (antiparallel) to the electron spin on the $\lambda = + (-)$ band at $\bm{k}$ as seen from Eqs.~\eqref{eq:spin_eval} and \eqref{eq:spin_eval_out}.

Instead of directly solving the linerized gap equation \eqref{eq:def_Kernel}, we introduce
\begin{equation} \label{eq:def_Kq}
    K(\bm{q}) = \frac{V}{\Omega} \sum_{\bm{k}} \sum_{\lambda,p}
    \frac{|u_p|^2}{4E^{(s)}_{\lambda,p}}
    \frac{\psi_{\bm{k}}^2\sinh{\beta E^{(s)}_{\lambda,p}}}{\cosh{\beta E^{(s)}_{\lambda,p}}+\cosh{\beta E^{(a)}_{\lambda,p}}}  - 1,
\end{equation}
and investigate $K_{\max} = \max_{\bm{q}} K(\bm{q}) = K(\bm{q}_{\max})$. 
Although we explicitly show only $\bm{q}$ as the argument of $K$, 
$K(\bm{q})$ and thus $K_{\max}$ depend on $\so$, $\AM$, and $T$. 
The other parameters such as $\mu$ are fixed as specified below.
The center-of-mass momentum which maximizes $K(\bm{q})$, denoted by $\bm{q}_{\max}$, is the momentum which is the most promising to be realized. However, far from the phase boundary, the optimal momentum should be determined by solving the gap equation self-consistently. The linearized gap equation can be represented as $K_{\max} = 0$ for the above-mentioned parameters. Particularly, we define $\Delta_{\mathrm{AM,c}}$ such that $K_{\max} = 0$ for given $\so$ and $T$. We also define $\bm{q}_{\max, \mathrm{c}}$ which maximizes $K(\bm{q})$ for $\Delta_{\mathrm{AM,c}}$.
As seen from the summation over $p$ in Eq.~\eqref{eq:def_Kq}, 
there are two contributions: one from the inter-band pairing and the other from the intra-band pairing.
Therefore, the dominant pairing channel can be estimated by comparing 
the weights of these two contributions.
\section{Results based on the linearized gap equation}
\begin{figure}[tb]
    \centering
    \includegraphics[width=0.8\linewidth]{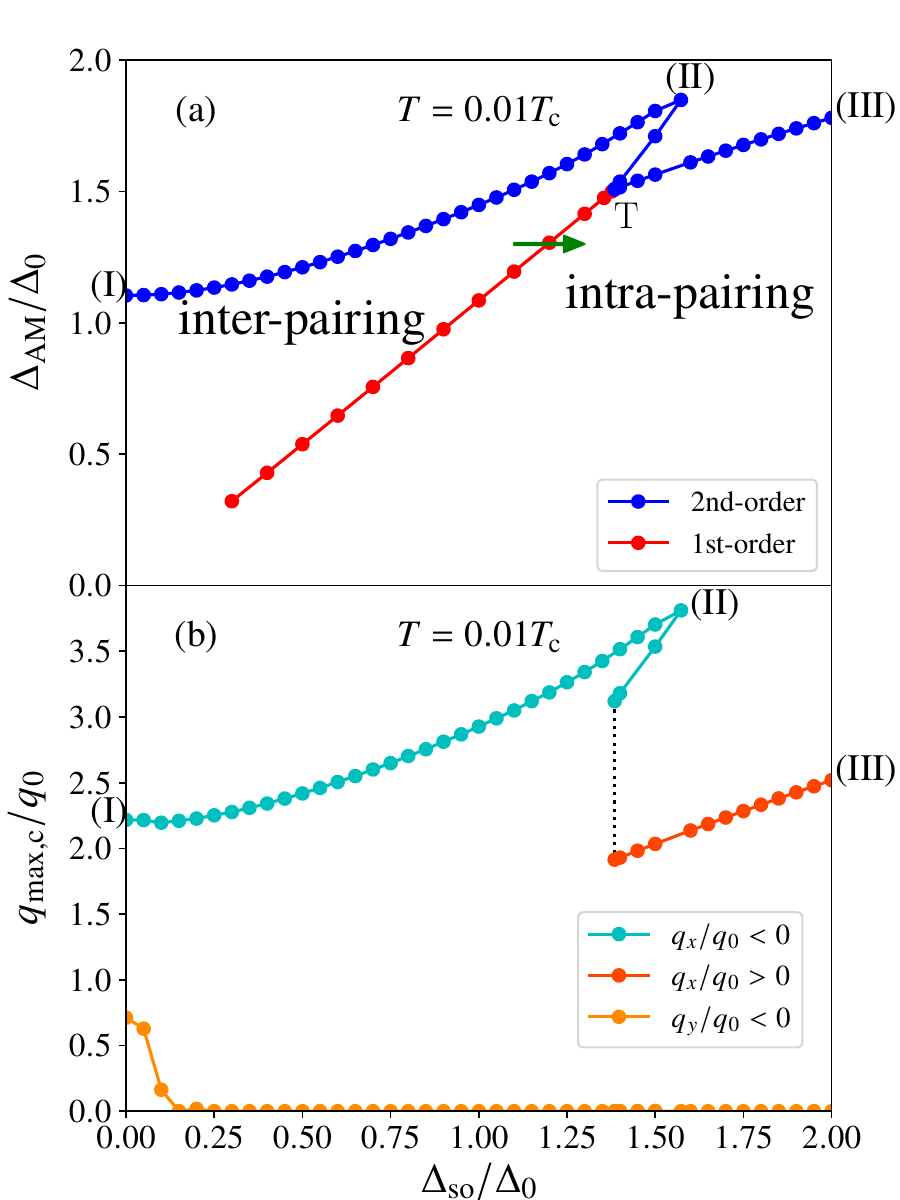}
    \caption{(a) Phase diagram for the in-plane altermagnet in the $\so$-$\AM$ plane at $T=0.01T_{\mathrm{c}}$. ``Inter-pairing'' (``intra-pairing'') denotes the superconducting states mainly from the inter-band pairing (the intra-band pairing). The blue line corresponds to $\Delta_{\mathrm{AM,c}}$ as a function of $\so$, which accounts for the phase boundary between the superconducting state and the normal state. The red line denotes the first-order transition line between the inter-band pairing and the intra-band pairing. (b) $\bm{q}_{\mathrm{max,c}}$ along the blue line in (a). The cyan (red) line shows negative (positive) $q_x$, $x$-component of $\bm{q}_{\mathrm{max,c}}$, and the orange line shows $q_y$. The dotted line represents a discontinuous change of $\bm{q}_{\mathrm{max,c}}$ at the tricritical point $\mathsf{T}$ in (a).}
    \label{fig:inplane_T=0.01}
\end{figure}
\subsection{In-plane altermanget}
\label{sec:inplane altermagnet}
We first present the phase diagram for the in-plane altermagnet in the $\so$-$\AM$ plane at a fixed temperature of $T=0.01T_{\mathrm{c}}$ in Fig.~\ref{fig:inplane_T=0.01}(a). We set $\bm{n}=\hat{\bm{y}}$ in Eq.~\eqref{eq:hvector} for simplicity. 
As units of the temperature, the energies, and the momentum of Cooper pairs,
we introduce the following quantities, respectively: $T_{\mathrm{c}}$, $\Delta_0$, and $q_0$, where 
$T_{\mathrm{c}}$ is the critical temperature, 
$\Delta_0=(\pi/e^{\gamma})k_{\mathrm{B}}T_{\mathrm{c}}$ is the $s$-wave superconducting gap at $T=0$, 
and $q_0 = \Delta_0/\varv_\mathrm{\mathrm{F}}$ 
with the Fermi velocity $\varv_{\mathrm{F}} = \sqrt{2\mu/m}$
corresponds to the inverse coherence-length, for $\so = \AM = 0$.
We set $\mu=50\Delta_0$ , and introduce $\omega_{\mathrm{D}}=10\Delta_0$ as a cutoff energy and calculate $K(\bm{q})$ in the $\bm{k}$ space satisfying $|\xi_k|<\omega_{\mathrm{D}}$.
The blue line in Fig.~\ref{fig:inplane_T=0.01}(a), along which $K_{\max} = 0$, accounts for the second-order transition between the superconducting state and the normal state. 
Below (Above) this line, $K_{\max} > 0$ ($K_{\max} < 0$), and thus the superconducting (normal) state is expected. 
The phase diagram consists of two superconducting states, both of which are the finite-momentum superconductivity. 
They are separated by the red line shown in Fig.~\ref{fig:inplane_T=0.01}(a), across which $\bm{q}_{\max}$ changes discontinuously. 
Thus, a first-order transition is expected, although this is based on the linearized analysis and requires the self-consistent calculation as shown in Sec.~\ref{sec:self-consistent_AM}.
The first-order transition line starts from the triple point $\mathsf{T}$ at $(\so^*/\Delta_0, \AM^*/\Delta_0)\approx(1.39, 1.51)$, which is on the second-order phase transition line between the two superconducting states and the normal state. 
We show the Cooper-pair momentum along the second-order phase transition line in Fig.~\ref{fig:inplane_T=0.01}(b), from which one can confirm the coexistence of the two different finite-momentum states as well as the normal state at $\mathsf{T}$. We also note that most of the phase boundary, the Cooper-pair momentum is along the $x$ direction, but only for small $\so$ region the $y$ component appears. The orientation of the Cooper-pair momentum is discussed in Appendix\ref{sec:orientation_of_q}, and in the following we focus on the case where the momentum is along the $x$ direction.

\begin{figure}[tb]
    \centering
    \includegraphics[width=1.0\linewidth]{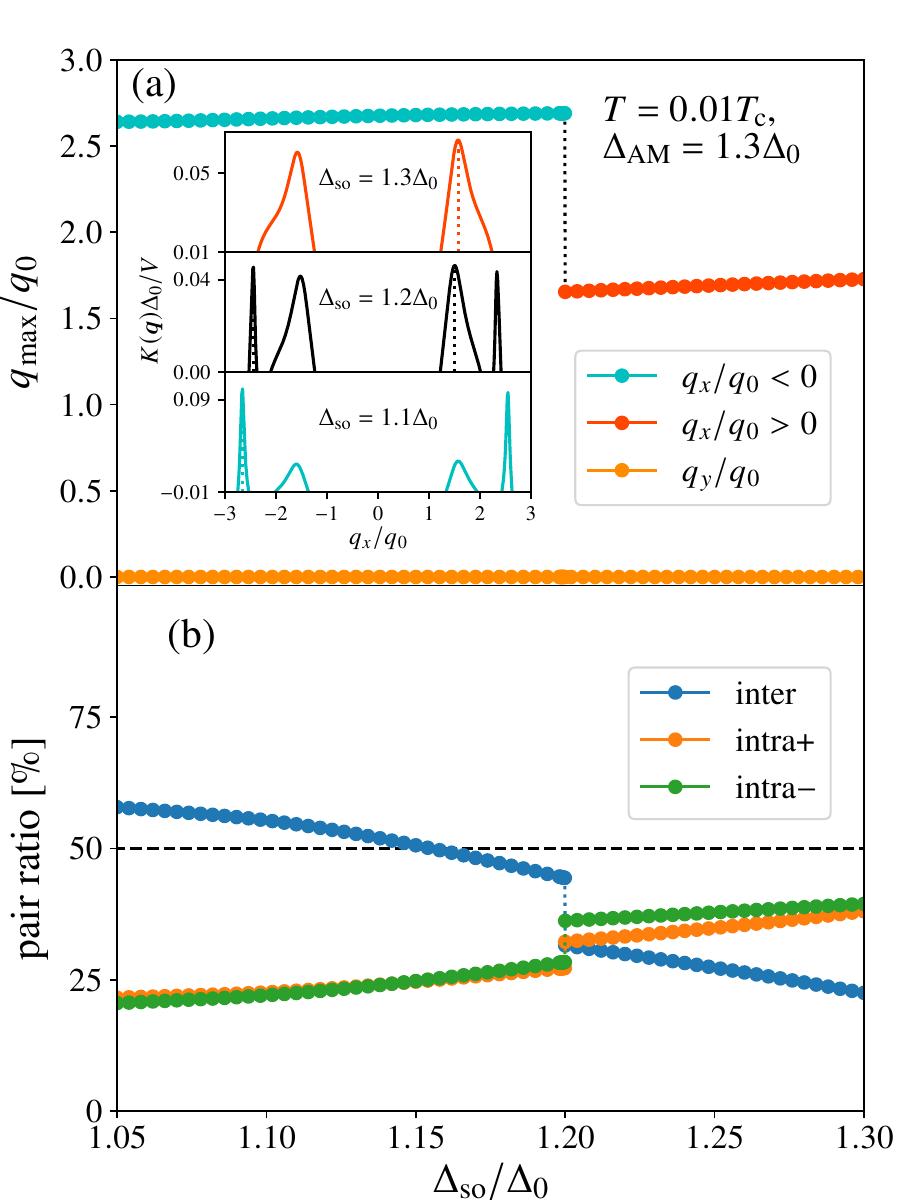}
    \caption{(a) $\bm{q}_{\mathrm{max}}$ and (b) contributions to $K(\bm{q}_{\mathrm{max}})$ from each pairing channel along the green arrow drawn in Fig.~\ref{fig:inplane_T=0.01}(a). $\AM$ is fixed to $1.3\Delta_0$, and $\so$ changes from $1.05\Delta_0$ to $1.3\Delta_0$. The inset in (a) shows the $q_x$ dependence of $K(\bm{q})$ for three $\so$'s with $q_y=0$. The dotted lines in each panel in the inset denote corresponding $\bm{q}_{\mathrm{max}}$. ``Intra$+$'' (``intra$-$'') in (b) represents the contributions from the intra-band pairing which consists of electrons on the $\lambda=+$ ($-$) band.}
    \label{fig:green-arrow}
\end{figure}

One can see that the first-order transition takes place when $\so$ and $\AM$ are comparable. 
Along the green arrow ($\AM = 1.3\Delta_0$) depicted in Fig.~\ref{fig:inplane_T=0.01}(a), we show the changes in the Cooper pair momentum $\bm{q}_{\max}$ and the pairing structure in Figs.~\ref{fig:green-arrow}(a) and \ref{fig:green-arrow}(b), respectively.
As seen from Fig.~\ref{fig:green-arrow}(a), 
the $y$-component of $\bm{q}_{\max}$ is zero in this range of $\so$, and the $x$-component changes its sign and magnitude when across the red line at $\so \sim 1.2\Delta_0$. 
Note that, when $\so$ is nonzero, $K(\bm{q})$ is not symmetric under the sign reversal of $q_x$ for $\bm{n} = \hat{\bm{y}}$ as shown in the inset of Fig.~\ref{fig:green-arrow}(a). 
In Fig.~\ref{fig:green-arrow}(b), the contribution to $K(\bm{q}_{\mathrm{max}})$ from each pairing channel is shown.
It is found that the inter-band pairing is dominant in the region $\so/\Delta_0\lesssim1.15$ while the intra-band pairing (the sum of ``intra$+$'' and ``intra$-$'') is dominant in the region $\so/\Delta_0\gtrsim1.15$. The contribution from each pairing displays a discontinuous change such that the intra-band pairing is more enhanced at $\so=1.2\Delta_0$,
where $\bm{q}_{\max}$ also changes discontinuously.
Considering that superconducting states far from the first-order transition line is well distinguished by the dominant pairing channel, we label the superconducting states above (below) the first-order transition line in Fig.~\ref{fig:inplane_T=0.01}(a) as the inter-band pairing (the intra-band pairing), although two pairing channels are comparable in the region $1.15\lesssim \so/\Delta_0\lesssim1.2$.
These results conclude that two different superconducting states correspond to the different pairing channels, accompanied by the discontinuous change of the optimal momentum of the Cooper pairs~\cite{helical_Zwicknagl_Jahns_Fulde_2017,topological-metal-liang-fu-2021}. 

\begin{figure}[tb]
    \centering
    \includegraphics[width=1.0\linewidth]{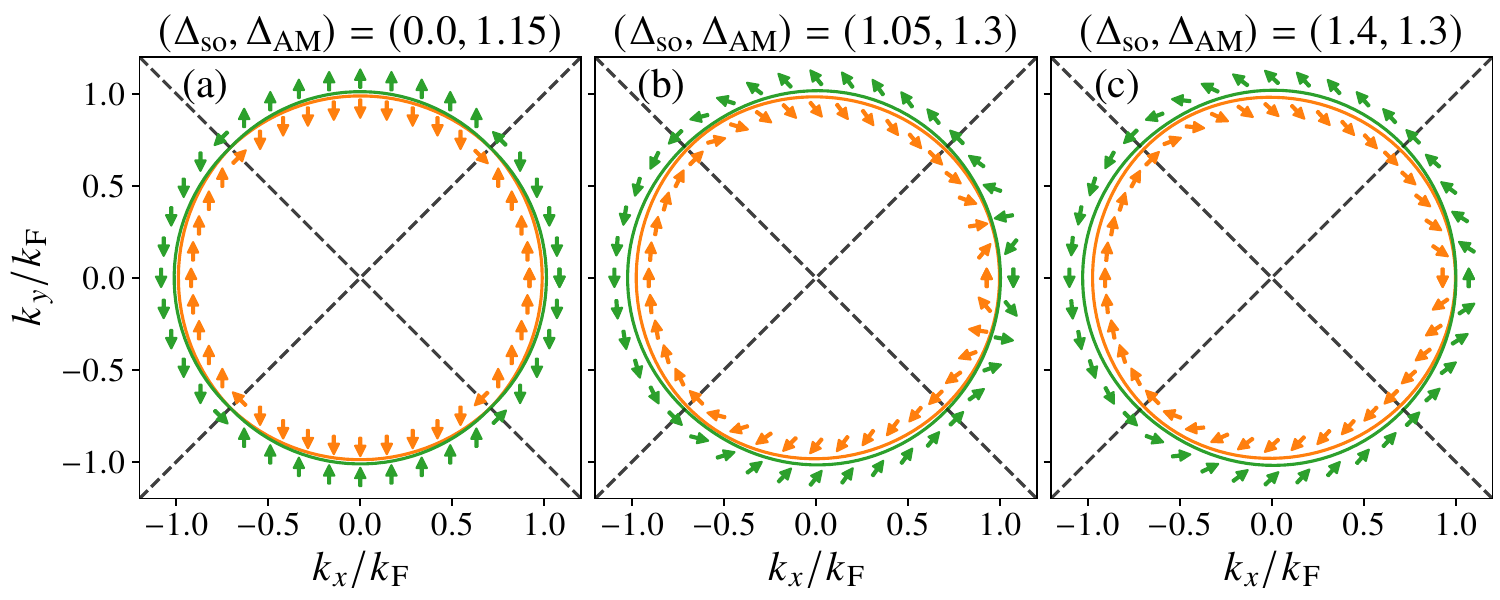}
    \caption{FSs ($E^{\mathrm{in}}_{\lambda}(\bm{k})=0$) and spin textures for (a) $\so=0$, (b) $\so<\AM$ and (c) $\so>\AM$. The orange (green) line represents the FS obtained from $E^{\mathrm{in}}_{\lambda}(\bm{k})=0$ for $\lambda=+$ ($-$).  Arrows indicate the direction of the in-plane spin polarization. The dotted lines correspond to the directions along which the altermagnetic spin-splitting is absent.}
    \label{fig:FS&spin}
\end{figure}

We explain the physical picture of the formation of the finite-momentum Cooper pair in terms of spin textures on FSs.
For the in-plane altermagnet, the energy dispersion [Eq.~\eqref{eq:dispersion}] is given by
\begin{equation}
    E^{\mathrm{in}}_{\lambda}(\bm{k})=\xi_k+\lambda\sqrt{\qty[\AM\qty(\hat{k}^2_x-\hat{k}^2_y)-\so\hat{k}_x]^2+\so^2\hat{k}_y^2},
\end{equation}
and the expectation value of the electron spin at momentum $\bm{k}$ on band $\lambda$ is
\begin{equation}\label{eq:spin_eval}
    \ev{\bm{\sigma}^{\mathrm{in}}_{\lambda}} = \frac{\lambda}{|\bm{h}(\bm{k})|}\qty(\so \hat{k}_y, \AM(\hat{k}^2_x-\hat{k}^2_y)-\so \hat{k}_x,0).
\end{equation}
The FSs of the normal states calculated from $E^{\mathrm{in}}_{\lambda}(\bm{k})=0$ for some parameter sets are shown in Fig. \ref{fig:FS&spin}. 
The orange (green) arrows show the in-plane spin polarization on the $\lambda=+(-)$ band. 
For $\so=0$ [Fig.~\ref{fig:FS&spin}(a)], there is ambiguity in the definition of the bands because the two bands are degenerated at the $\bm{k}$ points along $k_x=\pm k_y$, where the altermagnetic splitting is absent.
In our definition, the direction of the electron spin on each band for $\so=0$ is not constant but alternates every $90^{\circ} $ rotation for continuity to the case with nonzero $\so$.  
For $s$-wave superconductors, the conventional BCS-pairing can be formed by using electrons at degenerate points\cite{zero-field_Chakraborty_Black-Schaffer_2024}. 
However, such pairings are prohibited in the $d_{x^2-y^2}$-wave superconductors because the $d_{x^2-y^2}$-wave superconducting order parameter have the same node-directions as those for the altermagnetic spin-splitting. 
Other than the node directions, an up-spin electron at $(k_x,k_y)$ on one band cannot find a down-spin electron at $(-k_x,-k_y)$ on the same band, but can find it at $(-k_x+q_x,-k_y+q_y)$ on the other band. 
This mechanism makes the finite-momentum inter-band pairing stable~\cite{zero-field_Chakraborty_Black-Schaffer_2024,FMSC_in_AM_Sim_Knolle_2024}. 
A nonzero $\so$ lifts the degeneracies and induces anisotropic deformations of the two FSs, generating the spin texture in momentum space as seen from Eq.~\eqref{eq:spin_eval}.
We revisit the effects of the deformation later to investigate the intra-band pairing mechanism characteristic to the altermagnetic spin-splitting, and here focus on the spin texture.
For small $\so$ compared with $\AM$ [Fig.~\ref{fig:FS&spin}(b)], the altermagnetic spin configuration is dominant, and therefore, the inter-band pairing scenario is still applicable. 
For large $\so$ compared with $\AM$ [Fig.~\ref{fig:FS&spin}(c)], the spin directions are dominated by the RSOC, and thus the intra-band pairing becomes more stable than the inter-band pairing. 
We note that these two pairing channels have comparable contributions near the first-order transition line as mentioned above.

\begin{figure}
    \centering
    \includegraphics[width=\linewidth]{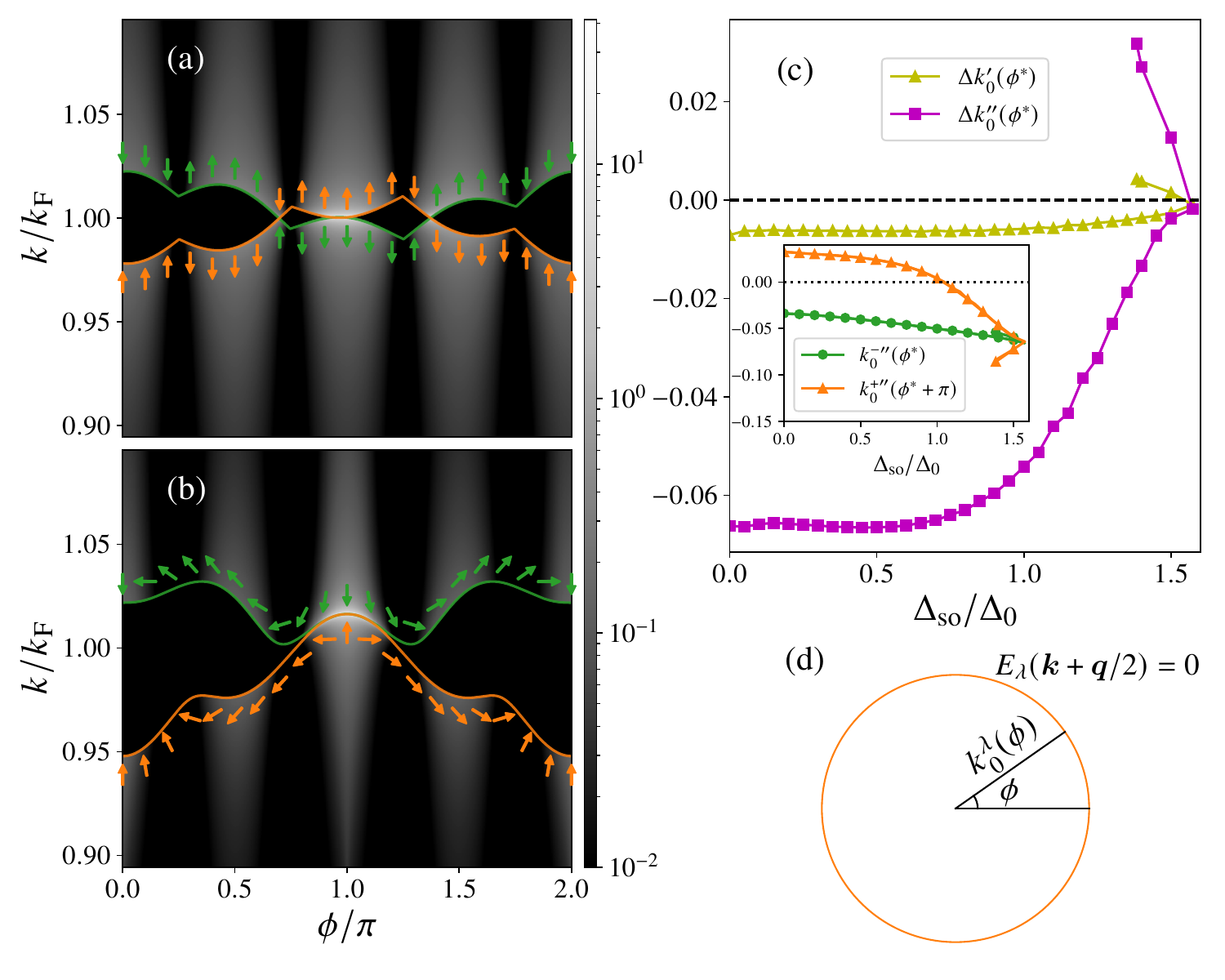}
    \caption{(a) (b) Two FSs $k^-_0(\phi)$ (green solid line) and $k^{+}_0(\phi+\pi)$ (orange solid line). The intensity of the integrand of $K(\bm{q})$ is also shown with color map. The parameter sets for each figure are explained in the main text. The arrows represent orientations of electron spins in the $k_x$-$k_y$ plane. (c)Values of the first derivative $\Delta k'_0(\phi^*)$, and the second derivative $\Delta k''_0(\phi^*)$ at the intersection point $\phi^*$ around $\pi$. The inset shows the second derivatives ${k^-_0}''(\phi^*)$ and ${k^+_0}''(\phi^*+\pi)$. (d)Schematic illustration showing the definition of $k^{\lambda}_0(\phi)$ and $\phi$. The orange line is an example of the FS calculated from $E_{\lambda}(\bm{k}+\bm{q}/2)=0$.}
    \label{fig:nesting_FS}
\end{figure}

In Fig.~\ref{fig:inplane_T=0.01}(a), the upturn of the second-order transition line in the inter-band pairing region and the emergence of the reentrant region are observed, implying that the finite-momentum superconductivity is stabilized at the larger $\so$ and $\AM$. The mechanism of the stabilization can be interpreted from the nesting condition for the FFLO state\cite{H_Shimahara_nesting}.
It has been reported that the enhancement of the critical field can be occurred when the two FSs $E_{\uparrow}(\bm{k}+\nobreak\bm{q}/2)=0$ and $E_{\downarrow}(-\bm{k}+\bm{q}/2)=0$ have a good fit.
To check the nesting condition for the present system, Figs.~\ref{fig:nesting_FS}(a) and \ref{fig:nesting_FS}(b) show the two FSs involved in the inter-band pairing. 
The green (orange) solid line represents $k^{\lambda=-}_0(\phi)$ [$k^{\lambda=+}_0(\phi+\pi)$], where $k^{\lambda}_0(\phi)$ is the radial length of the FS as a function of the azimuthal angle $\phi$ and defined as
\begin{equation} \label{eq:def_k0}
E^{\mathrm{in}}_{\lambda}\qty(k^{\lambda}_0(\phi)\cos{\phi}+q_x/2,\,k^{\lambda}_0(\phi)\sin{\phi}+q_y/2)=0.
\end{equation}
The intensity of the integrand of $K(\bm{q})$ at given parameters is also shown with color map.
We choose two sets of parameters, (I) $(\so/\Delta_0, \AM/\Delta_0, q_x/q_0, q_y/q_0)\approx(0.0, 1.10, -2.22, -0.711)$ for panel (a), and (II) $(\so/\Delta_0, \AM/\Delta_0, q_x/q_0, q_y/q_0)\approx(1.57, 1.85, -3.81, 0.0)$ for panel (b), which are also indicated in Figs.~\ref{fig:inplane_T=0.01}(a) and \ref{fig:inplane_T=0.01}(b) by (I) and (II), respectively.
As can be seen from the color map, the intensity of the integrand is enhanced when the two FSs cross or touch with each other in the area other than the nodes of the $d$-wave superconducting order parameter. 
By comparing Figs.~\ref{fig:nesting_FS}(a) and \ref{fig:nesting_FS}(b), one can see the two FSs in Fig.~\ref{fig:nesting_FS}(b) have a better fit around $\phi=\pi$. To evaluate how the two FSs touch with each other, we introduce $\Delta k_0(\phi)=k^{-}_0(\phi)-k^{+}_0(\phi+\pi)$ representing the difference of two radii involved in the formation of Cooper pairs. 
Figure~\ref{fig:nesting_FS}(c) shows the values of the first derivative $\Delta k'_0(\phi^*)$, and the second derivative $\Delta k''_0(\phi^*)$ at an intersection point $\phi^*$ of the two FSs, $\Delta k_0(\phi^*) = 0$, around $\phi = \pi$, for the parameter sets along the second-order transition line of the inter-band pairing shown in Fig.~\ref{fig:inplane_T=0.01}(a).
Note that nonzero $\Delta k'_0(\phi^*)$ accounts for the crossing of $k_0^{-}(\phi)$ and $k_0^+(\phi + \pi)$ at $\phi =\phi^*$. 
One can see that as $\so$ increases,  $|\Delta k'_0(\phi^*)|$ goes to zero and the two FSs touch at $\phi = \phi^*$. At the same time, $|\Delta k''_0(\phi^*)|$ also goes to zero.
By focusing on each second derivative $k_0^{\lambda \prime\prime}$ shown in the inset of Fig.~\ref{fig:nesting_FS}(c), one can see that ${k^+_0}''(\phi+\pi)$ exhibits the sign change at $\so\sim\Delta_0$, and the two FSs come to have the same curvature, that is, the better fit of the FSs is realized, as shown in Fig.~\ref{fig:nesting_FS}(b).
Therefore, the well-nested FSs stabilize the inter-band finite-momentum superconductivity with large $\so$.
We note that the good nesting condition and the upturn of the critical magnetic field are not observed in the case for the uniform magnetic field as shown in Appendix~\ref{sec:uniform magnetic field}.

\begin{figure}[tb]
    \centering
    \includegraphics[width=\linewidth]{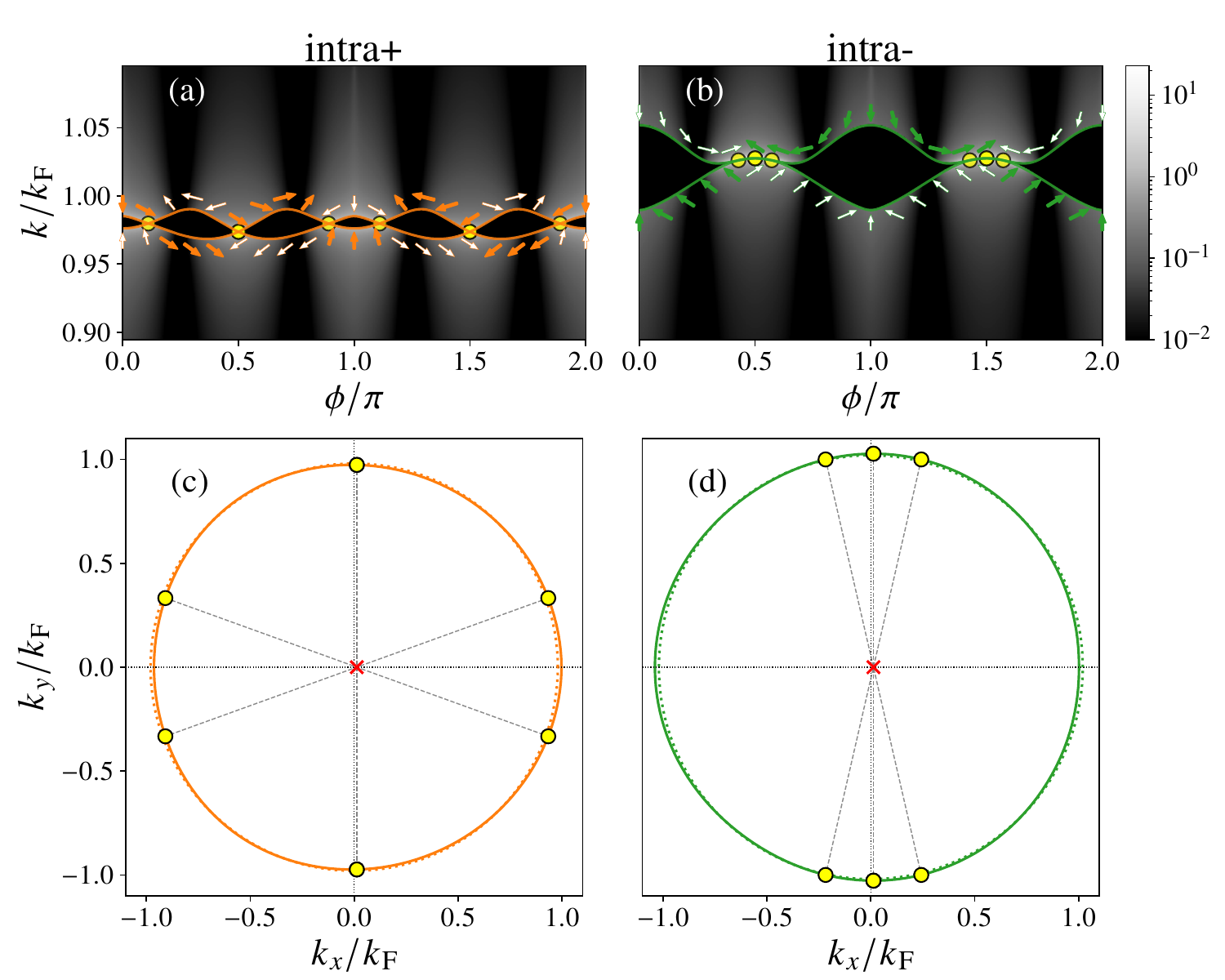}
    \caption{Illustrations of the intra-band pairing on the FSs for $(\so/\Delta_0, \AM/\Delta_0,q_x/q_0)=(2.0,1.78, 2.52)$. $k^{\lambda}(\phi)$ and $k^{\lambda}(\phi+\pi)$ are shown in (a) with orange lines ($\lambda=+$) and in (b) with green lines ($\lambda=-$). The intensity of the integrand of $K(\bm{q})$ is shown with the color map, and $\braket{\bm{\sigma}_{\lambda}^{\mathrm{in}}}$ along the FSs is shown by the filled allows for $k^{\lambda}(\phi)$ and the open arrows for $k^{\lambda}(\phi+\pi)$ . The FSs calculated from $E^{\lambda}(\bm{k})=0$ are shown in (c) ($\lambda=+$) and in (d) ($\lambda=-$) with solid lines. The FSs with $\AM=0$ are also shown for comparison with dotted lines. Yellow circles in (a, b) indicate intersections of $k^{\lambda}(\phi)$ and $k^{\lambda}(\phi+\pi)$. They correspond to the electrons to form Cooper pairs depicted by yellow circles in (c, d). The red cross mark in each bottom panel indicates the center-of-mass momentum of Cooper pairs.}
    \label{fig:intra_nesting_AM}
\end{figure}
Another noticeable feature in the altermagnet with the RSOC is that the intra-band pairing with a positive momentum is favored in the wide range of parameter.
In an $s$-wave superconductor with the RSOC and the Zeeman coupling, which we call the Rashba--Zeeman superconductor,
the outer FS always has larger density of states than one of the inner FS.
When a magnetic field is applied along the $y$ direction, the outer FS is shifted in $-k_x$ direction.
Thus, the intra-band pairing in the outer FS with a negative Cooper-pair momentum is favored in a Rashba--Zeeman superconductor\cite{daido_diode, PhysRevB-daido-diode, Smidman_Salamon_Yuan_Agterberg_2017, Yanase_Sigrist_2008}.
The superconducting state with a positive momentum is less favored because of the small density of states of the inner FS.
However, in the case of the in-plane altermagnet with the RSOC, although the situation of the density of states is the same as in the Rashba--Zeeman superconductors, the Cooper pair momentum becomes positive.
As seen in Fig.~\ref{fig:green-arrow}(b), in the intra-band pairing phase, both of inner and outer FSs almost equally contribute to the promising pairing, in contrast to the aforementioned situation in the Rashba--Zeeman superconductors that the intra-band pairing can be stabilized only in one FS.
The reason of the contribution can be understood from the deformation of the FSs.
Importantly, the FSs affected by the in-plane N\'{e}el vector and the RSOC do not display a shift of the whole FSs observed in the Rashba--Zeeman superconductors because of the node directions of the altermagnetic spin-splitting.
Instead, they display the anisotropic deformation depending on the azimuthal angle $\phi$.
We explain this deformation with the example shown in Fig.~\ref{fig:intra_nesting_AM}.
The FSs for the intra-band pairing are shown in Fig.~\ref{fig:intra_nesting_AM} for the parameter set at the second-order transition point with $\so=2.0\Delta_0$ indicated by (III) in Figs.~\ref{fig:inplane_T=0.01}(a) and~\ref{fig:inplane_T=0.01}(b).
Focusing on the outer FS in Fig.~\ref{fig:intra_nesting_AM}(d), one can see that it exhibits a shift toward the $-k_x$ ($+k_x$) direction in $-\pi/4+m\pi\le\phi\le\pi/4+m\pi$ ($\pi/4+m\pi\le\phi\le3\pi/4+m\pi$) ($m=0,1$).
The inner FS also shows such an anisotropic deformation, whose sign is opposite to that of the outer FS.[Fig.~\ref{fig:intra_nesting_AM}(c)].
Such a deformation implies that one FS can contribute to the positive- and negative-Cooper-pair-momentum states because the shifts in both directions coexist in one FS. In other words, both of the FSs can contribute to the finite momentum state along each direction.
Which direction of the momentum is optimized is determined by the density of states and the nesting condition.
Using Fig.~\ref{fig:intra_nesting_AM} as examples, the positive momentum is selected, resulting in a good fit around $\phi=0.5\pi,1.5\pi$ in the outer FS [\ref{fig:intra_nesting_AM}(a)] and in Cooper pairs around $\phi=0,\pi$ in the inner FS [\ref{fig:intra_nesting_AM}(b)]. 
The other situation corresponding to the negative momentum, where the roles of the outer and inner FSs are opposite, is realized in SC2 obtained by the self-consistent calculation shown in Sec.~\ref{sec:self-consistent_AM}. 
This stabilization mechanism is applicable for other parameters for $\so>\AM$.
Compared to the case of the uniform magnetic field in Fig.~\ref{fig:Zeeman_nesting}, it can be found that the momentum-dependent splitting due to the in-plane N\'{e}el vector is a key to cause the above-mentioned mechanism of the intra-band pairing.

\begin{figure}[tb]
    \centering
    \includegraphics[width=0.8\linewidth]{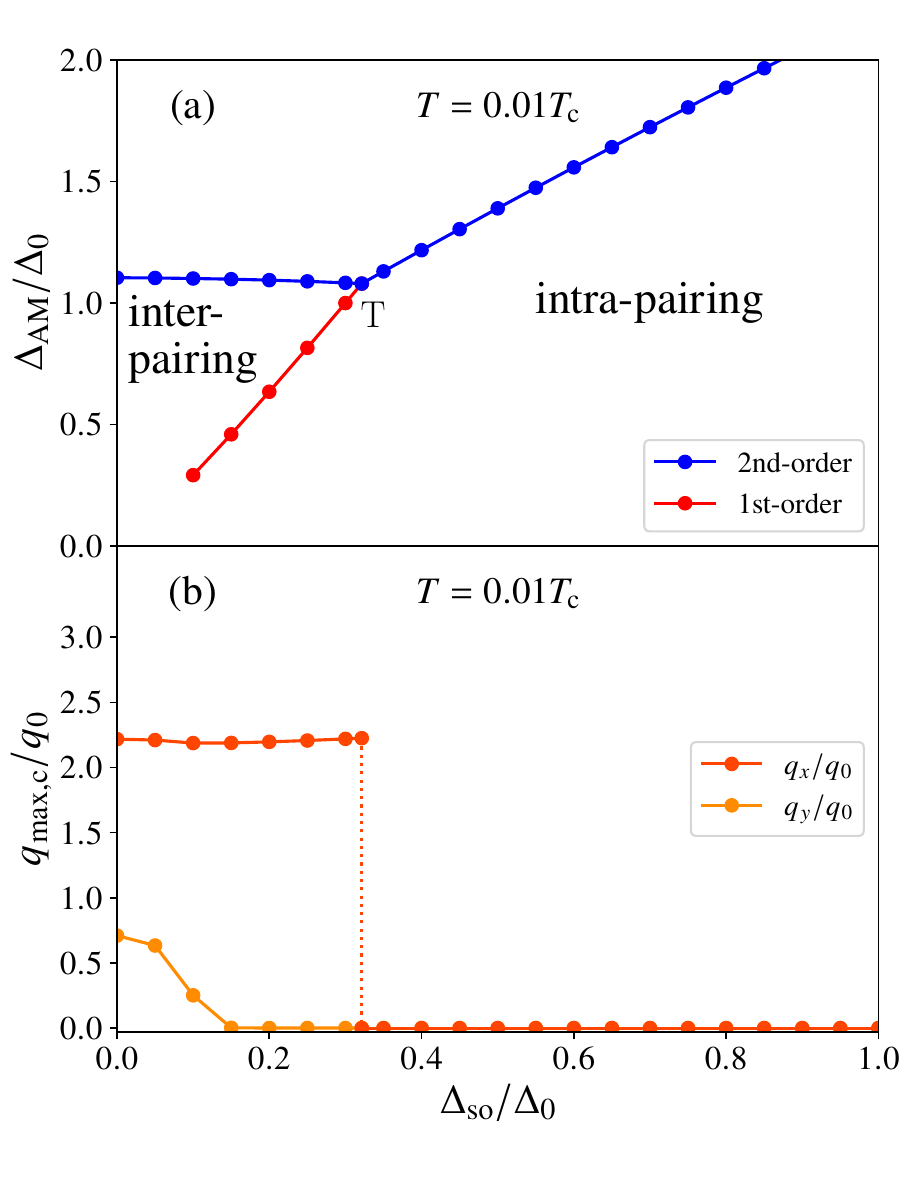}
    \caption{(a) Phase diagram of the out-of-plane altermagnet in the $\so$-$\AM$ plane. (b) $\bm{q}_{\mathrm{max,c}}$ along the second-order transition line in (a). Dotted line represents an discontinuous change of $\bm{q}_{\mathrm{max,c}}$ at the triple point. Definitions of the lines are the same as those in Fig. \ref{fig:inplane_T=0.01}.}
    \label{fig:outofplane_T=0.01}
\end{figure}
\subsection{Out-of-plane altermagnet}
\begin{figure}[tb]
    \centering
    \includegraphics[width=1.0\linewidth]{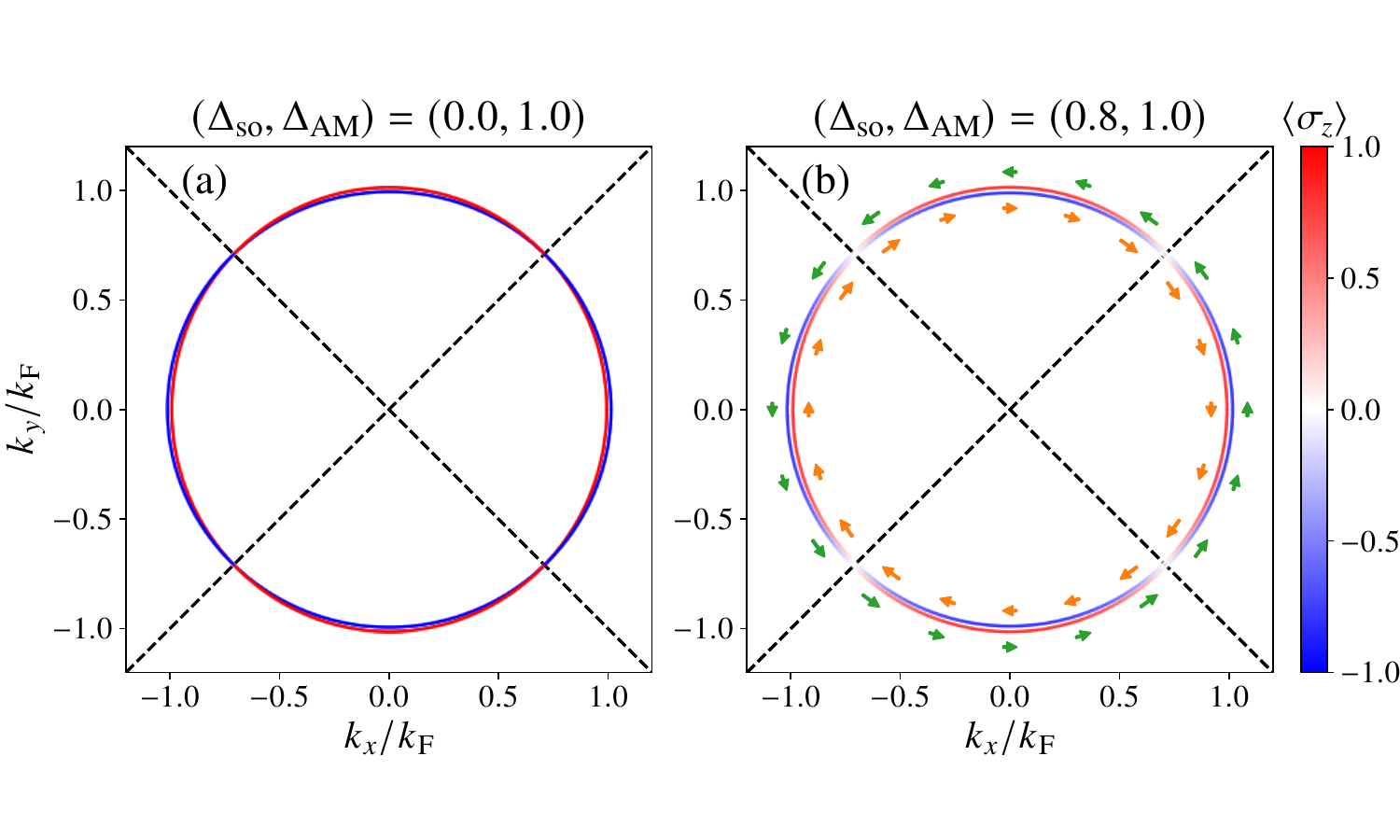}
    \caption{FSs and spin textures for (a) $\so=0$ case and (b) $\so\ne0$ case. The color of FSs indicates the magnitude of $\ev{\sigma_z}$, the out-of-plane spin polarization. Color arrows indicate the direction of the in-plane spin polarization. Along the dotted lines, the altermagnetic spin-splitting is zero.}
    \label{fig:outofplaneFS&spin}
\end{figure}
Figure~\ref{fig:outofplane_T=0.01} shows the $\so$-$\AM$ phase diagram at $T=0.01T_{\mathrm{c}}$ for the out-of-plane altermagnet which is described by $\bm{n} = \hat{\bm{z}}$ in Eq.~\eqref{eq:hvector}. 
Note that $K(\bm{q})$ is symmetric under the sign reversal of $\bm{q}$ in the out-of-plane case, but we consider the single-$\bm{q}$ finite-momentum superconductivity, which corresponds to the FF state.
The phase diagram also consists of the inter-band pairing superconductivity (above the red line) and the intra-band pairing superconductivity (below of the red line). 
The Cooper-pair momentum is finite only for the inter-band pairing, and the BCS-pairing is stable for the intra-band pairing, in contrast to the case of the in-plane N\'{e}el vector. 
The triple point $\mathsf{T}$ in the out-of-plane case is $(\so^*/\Delta_0, \AM^*/\Delta_0)\approx(0.32, 1.08)$, and it is shifted to the lower $\so$ side compared with the one in the in-plane case as discussed later. 
Consequently, compared to the in-plane altermagnet, the finite-momentum superconductivity is realized in a narrow region of the phase diagram.

The energy dispersion in the out-of-plane altermagnet can be obtained from Eq.~\eqref{eq:dispersion} as
\begin{equation}
    E^{\mathrm{out}}_{\lambda}(\bm{k}) = \xi_k +\lambda\sqrt{\so^2\hat{k}^2+\AM^2(\hat{k}_x^2-\hat{k}_y^2)^2}.
\end{equation}
The expectation value of the electron spin at momentum $\bm{k}$ on band $\lambda$ is given by
\begin{equation}\label{eq:spin_eval_out}
    \ev{\bm{\sigma}^{\mathrm{out}}_{\lambda}} = \frac{\lambda}{|\bm{h}(\bm{k})|}\qty(\so \hat{k}_y, -\so \hat{k}_x,\AM(\hat{k}^2_x-\hat{k}^2_y)).
\end{equation}
FSs calculated from $E^{\mathrm{out}}_{\lambda}(\bm{k})=0$ and the corresponding spin textures are shown in Fig. \ref{fig:outofplaneFS&spin}.
In the in-plane altermagnet, the RSOC and the altermagnetic spin-splitting are coupled as can be seen in the $y$-component in Eq.~\eqref{eq:spin_eval}.  
However, they are decoupled in the out-of-plane altermagnet as can be seen in Eq.~\eqref{eq:spin_eval_out}. Thus, a deformation and a shift of the FSs similar to those observed in the in-plane altermagnet are absent in the out-of-plane altermagnet. As for electron spins, while the RSOC affects the in-plane spin polarization, the altermagnetic spin-splitting affects the out-of-plane spin polarization.
Such decoupling of the N\'{e}el vector and the RSOC makes different properties of the inter-band pairing and the intra-band pairing as follows.
For the case without the RSOC, electron spins have no in-plane components [Fig. \ref{fig:outofplaneFS&spin}(a)] and the finite-momentum inter-band pairing caused by the altermagnetic spin-splitting is realized.
This scenario is the same as that of the in-plane altermagnet without the RSOC, and applicable for small $\so$. 
For the case with the RSOC, the in-plane components of electron spins become finite, and the in-plane component of the spin at $\bm{k}$ on one band is always antiparallel to that at $-\bm{k}$ on the same band, as seen from Eq.~\eqref{eq:spin_eval_out}.
Thus, the intra-band pairing with finite $\so$ makes the BCS-pairing stable instead of the finite-momentum pairing.

As mentioned above, the triple point and the first-order transition line are shifted toward the smaller $\so$ region. This result implies that the effect of the RSOC is detrimental to the finite-momentum inter-band pairing in the out-of-plane altermagnet.
This is because almost all the region of the FSs other than the nodal directions of the $d$-wave superconducting order parameter can contribute to the intra-band BCS pairing, in contrast to the case of the finite-momentum inter-band pairing as demonstrated in Fig.~\ref{fig:nesting_FS}(a).
Thus, we can say that the shrink of the inter-band pairing region results from the difference in the electron population to contribute to the inter- and intra-band pairing.

\begin{figure*}[tb]
    \centering
    \includegraphics[width=\linewidth]{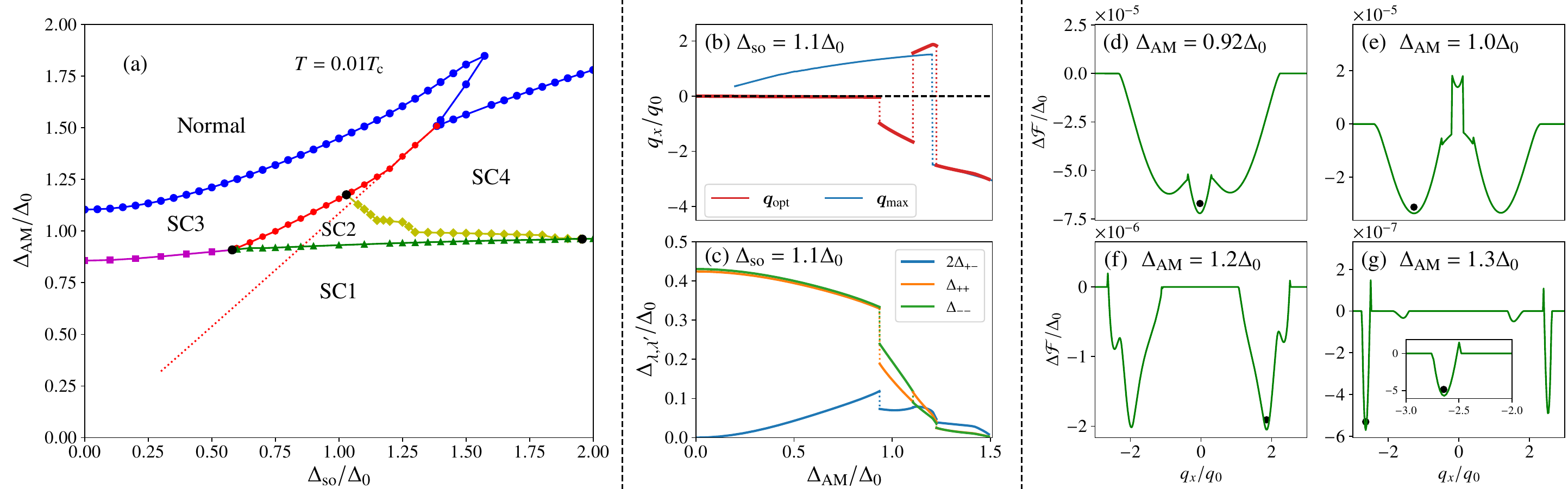}
    \caption{(a) Phase diagram obtained by the self-consistent calculation. The blue line is the second-order transition line between superconducting states and normal states. The superconducting states can be distinguished by the momentum $\bm{q}_{\mathrm{opt}}$ and their phase boundaries are shown by color lines with different markers. The dotted line is the first-order transition line shown in Fig.~\ref{fig:inplane_T=0.01}. (b) $\AM$ dependence of the momentum for $\so=1.1\Delta_0$. The red (blue) lines show $\bm{q}_{\mathrm{opt}}$ ($\bm{q}_{\mathrm{max}}$).  The dotted lines denote the discontinuous change caused by the phase transition. (c) $\AM$ dependence of $\Delta_{\lambda,{\lambda^{'}}}(\bm{q}_{\mathrm{opt}})$ for $\so=1.1\Delta_0$. (d-g) Condensation energies for several values of $\AM$ with $\so = 1.1\Delta_0$. $\bm{q}_{\mathrm{opt}}$ is depicted by black circle.}
    \label{fig:PD_BdG}
\end{figure*}
\section{Results based on self-consistent calculation}\label{sec:self-consistent_AM}
In this section, we further investigate the superconducting region in the phase diagram shown in Fig.~\ref{fig:inplane_T=0.01}(a) within the self-consistent calculation which is applicable also in a region far from the phase boundary.
The mean-field Hamiltonian in Eq.~\eqref{eq:MF_Hamiltonian} reads 
 \begin{equation}
    \begin{aligned}
         \hat{H}^{\mathrm{MF}} &= \frac{1}{2}\sum_{\bm{k}}\hat{\bm{C}}^{\dagger}(\bm{k},\bm{q})\check{H}_{\mathrm{BdG}}(\bm{k},\bm{q})\hat{\bm{C}}(\bm{k},\bm{q})\\
         &\quad\quad + \frac{1}{2}\sum_{\bm{k}\sigma}[H_0(\bm{k})]_{\sigma\sigma}+\frac{\Omega}{V}|\Delta(\bm{q})|^2
     \end{aligned}
 \end{equation}
with
 \begin{equation}
    \hat{\bm{C}}^{\dagger}(\bm{k},\,\bm{q})=\qty(\hat{c}^{\dagger}_{\bm{k}_+\uparrow},\,\hat{c}^{\dagger}_{\bm{k}_+\downarrow},\,\hat{c}_{-\bm{k}_-\uparrow},\,\hat{c}_{-\bm{k}_-\downarrow}).
 \end{equation}
The eigenstates and eigenvalues of $\check{H}_{\mathrm{BdG}}(\bm{k},\bm{q})$ are given by
\begin{equation}
    \check{H}_{\mathrm{BdG}}(\bm{k},\bm{q})
    \begin{pmatrix}
    \bm{u}_n(\bm{k},\bm{q})
    \\    
    \bm{v}_n(\bm{k},\bm{q})
    \end{pmatrix}=\varepsilon_{n}(\bm{k},\bm{q})
    \begin{pmatrix}
    \bm{u}_n(\bm{k},\bm{q})
    \\    
    \bm{v}_n(\bm{k},\bm{q})
    \end{pmatrix}.
\end{equation}
By using the two-component vectors $\bm{u}_n(\bm{k},\bm{q})$ and $\bm{v}_n(\bm{k},\bm{q})$, and $\varepsilon_{n}(\bm{k},\bm{q})$, the gap equation \eqref{eq:def_Delta} with Eqs.~\eqref{eq:Delta} and \eqref{eq:Vk} can be rewritten as
\begin{equation} \label{eq:gapeq_SCF}
    \Delta(\bm{q}) = - \frac{V}{\Omega}\sum_{\bm{k}n}\psi_{\bm{k}}[\bm{u}_n(\bm{k},\bm{q})]_\uparrow[\bm{v}_n(\bm{k},\bm{q})]^*_\downarrow f(\varepsilon_n(\bm{k}, \bm{q})).
\end{equation}
The gap function in Eq.~\eqref{eq:gapeq_SCF} can be decomposed into the band-based gap function as follows.
Using the relation between $\hat{c}_{\bm{k}\sigma}$ and $\hat{b}_{\bm{k}\lambda}$ introduced in Sec.~\ref{sec:model and method}, $\Delta(\bm{q})$ can be written as 
$\Delta(\bm{q})
=\sum_{\lambda,\lambda^{'}} \Delta_{\lambda,\lambda^{'}}(\bm{q}),
$
where $\Delta_{\lambda,\lambda^{'}}(\bm{q})$
is the band-based gap function given as
\begin{align}
    \Delta_{\lambda,\lambda^{'}}(\bm{q})=
    -\dfrac{V}{\Omega}\sum_{\bm{k}}\psi_{\bm{k}}[\bm{w}_{\lambda}(\bm{k}_+)]_{\uparrow} \braket{\hat{b}_{\bm{k}_+\lambda}\hat{b}_{-\bm{k}_-\lambda'}} [\bm{w}_{\lambda'}(-\bm{k}_-)]_{\downarrow}
\end{align}
with
\begin{align}
    \braket{\hat{b}_{\bm{k}_+\lambda}\hat{b}_{-\bm{k}_-\lambda'}}
    &=\sum_n f(\varepsilon_n(\bm{k},\bm{q}))~[\bm{w}_{\lambda}(\bm{k}_+)^\dagger
    \cdot \bm{u}_n(\bm{k},\bm{q})] 
    \nonumber \\
    &\hspace{2em}\times 
     [\bm{w}_{\lambda'}(-\bm{k}_-)^\dagger
    \cdot \bm{v}_n(\bm{k},\bm{q})^*]
\end{align}
This band-based gap function is used to analyze the contributions to the total gap, as performed in Fig.~\ref{fig:green-arrow}(b) even in the non-linear regime.

The free energy density $F_s(\Delta(\bm{q}), \bm{q})$ is calculated from the mean-field Hamiltonian and given by
\begin{equation}
    \begin{aligned}
        F_s(\Delta(\bm{q}), \bm{q}) &= \frac{1}{2\Omega}\sum_{\bm{k}\sigma}[H_0(\bm{k})]_{\sigma\sigma}+\frac{|\Delta(\bm{q})|^2}{V} \\
        &\hspace{1em}-\frac{k_{\mathrm{B}}T}{2\Omega}\sum_{\bm{k}n}\ln\qty(1+e^{-\varepsilon_n(\bm{k},\bm{q})/k_{\mathrm{B}}T}).
    \end{aligned}
\end{equation}
To discuss the stable superconducting states, we introduce the condensation energy $\Delta\mathcal{F}(\bm{q})$ by
\begin{equation}
    \Delta\mathcal{F}(\bm{q}) = F_s(\Delta(\bm{q}), \bm{q}) - F_s(0, \bm{q}).
\end{equation}
For a fixed $\bm{q}$, we solve Eq.~\eqref{eq:gapeq_SCF} self-consistently, and obtain $\Delta \mathcal{F}(\bm{q})$. The stable state, whose Cooper pair momentum is denoted by $\bm{q}_{\mathrm{opt}}$,
is determined 
so that $\Delta \mathcal{F}(\bm{q}_{\mathrm{opt}}) = \min_{\bm{q}} \Delta\mathcal{F}(\bm{q})$. 
In the following calculation, we set $\bm{q}=(q_x, 0)$ for simplicity, though this simplification is not valid for small $\so$.

For the in-plane altermagnet, we show the self-consistently calculated phase diagram  in the $\so$-$\AM$ plane at the fixed temperature $T=0.01T_{\mathrm{c}}$ in Fig.~\ref{fig:PD_BdG}(a). 
The phase diagram consists of four superconducting regions (SC1-SC4), which are distinguished by the momentum $\bm{q}_{\mathrm{opt}}$. The phase transitions between these states are first-order as shown by the red, purple, and green lines, and the black circles represent the triple points.
To see the characteristics of each phase, the $\AM$ dependence of $\bm{q}_{\mathrm{opt}}$ and $\Delta_{\lambda,\lambda^{'}}(\bm{q}_{\mathrm{opt}})$ for $\so=1.1\Delta_0$ are shown in Figs.~\ref{fig:PD_BdG}(b) and~\ref{fig:PD_BdG}(c), respectively. 
The momentum $\bm{q}_{\mathrm{max}}$ calculated from $K(\bm{q})$ is also shown in each panel for comparison. Landscapes of the condensation energy corresponding to these phases (SC1-SC4) are shown in Figs.~\ref{fig:PD_BdG}(d)-\ref{fig:PD_BdG}(g).

For $\so=1.1\Delta_0$ [Fig.~\ref{fig:PD_BdG}(b)], in the region $0<\AM/\Delta_0\lesssim0.94$,~the momentum is finite but very small, and its magnitude slightly increases with $\AM$ (SC1).
At $\AM \simeq 0.94\Delta_0$, there is a discontinuous transition to a large momentum superconducting state (SC2).
In the region $0.94\lesssim\AM/\Delta_0\lesssim1.22$, there are two phases (SC2 and SC4).
Their magnitude of the momentum is almost the same although their directions are opposite. 
With increasing $\AM$, both of the magnitude and the direction of the momentum again displays the discontinuous change between SC4 and SC3. Compared with the results based on the linearized gap equation, momentum changes corresponding to SC1 and SC2 are unique to the results by the self-consistent calculation.

Here, we discuss the pairing mechanism of each superconducting phase in terms of $\Delta_{\lambda,\lambda^{'}}(\bm{q}_{\mathrm{opt}})$ shown in Fig.~\ref{fig:PD_BdG}(c).
It is observed that SC1 consists of almost equal contributions from both $\Delta_{++}$ and $\Delta_{--}$. We have confirmed that the relative importance of the intra-band pairing ($\Delta_{++}+ \Delta_{--}$) and the inter-band pairing ($2\Delta_{+-}$) depends on $\so$. 
Considering that SC1 is realized in the small $\AM$ region with the small momentum, we recognize SC1 resembles the BCS-pairing [see also Figs.~\ref{fig:PD_BdG}(b) and \ref{fig:PD_BdG}(d)].
With regard to SC2 and SC4, they are dominated by $\Delta_{++}+\Delta_{--}$, 
thus, both of them are the intra-band pairing.
The most dominant component is $\Delta_{--}$ ($\Delta_{++}$) in SC2 (SC4), and the difference on the dominant pairing between SC2 and SC4 is manifested in the $\bm{k}$-resolved intensity, the integrand of $\Delta_{\lambda,\lambda}(\bm{q}_{\mathrm{opt}})$ for $\lambda = \pm$.
The $\bm{k}$-resolved intensity of $\Delta_{\lambda,\lambda}(\bm{q}_{\mathrm{opt}})$ for SC4 is similar to that of $K(\bm{q})$ in Figs.~\ref{fig:intra_nesting_AM}(a) (for $\lambda = -$) and~\ref{fig:intra_nesting_AM}(b) (for $\lambda = +$), while that for SC2 corresponds to the case with a negative momentum mentioned in the end of Sec.\ref{sec:inplane altermagnet}.
We remark that SC2 and SC4 are energetically competitive. Figure~\ref{fig:PD_BdG}(e) [\ref{fig:PD_BdG}(f)] shows the condensation energy landscape in the case that SC2 (SC4) is the most stable.
Both of them have large minimum-structures both in the positive and negative $q_x$ regions, and the minimum in the positive $q_x$ region in Fig.~\ref{fig:PD_BdG}(e) [in the negative $q_x$ region in Fig.~\ref{fig:PD_BdG}(f)] corresponds to SC4 (SC2) as a metastable state, whose energy is slightly larger than that of the most stable state.
Thus, in this parameter region, although either SC2 or SC4 is energetically favored, their slight difference of the energy cannot exclude the possibility of the two momenta state such as the LO state and the stripe phase,SC3 is dominated by $2\Delta_{+-}$ except in the vicinity of the phase transition between SC4 and SC3, implying that SC3 is the inter-band pairing. 
In addition, as can be seen in Fig.~\ref{fig:PD_BdG}(a), the first-order transition line between SC3 and SC4 in the vicinity of the second-order phase transition (red solid line) coincides with the one by the linearized gap equation (red dotted line). 
This guarantees the validity of the analysis by $K(\bm{q})$ in this regions. 
 Therefore, the pairing mechanism in Fig.~\ref{fig:green-arrow} can also confirm that SC3 (SC4) corresponds to the inter-band (intra-band) pairing.

We briefly mention a large $\so$ region, which is not shown in Fig.~\ref{fig:PD_BdG}(a). The first-order transition line between SC1 and SC4 does not terminate at a critical point, in contrast to the one between the ``$|\bm{q}|\sim 0$'' state and the ``intra-band pairing'' state in Fig.~\ref{fig:PD_Zeeman}. This is attributed to the difference in the sign of the Cooper-pair momenta between SC1 and SC4. However, we have confirmed that there is an additional finite-momentum superconducting state, which we call SC5, in a large $\so$ region, sharing the sign of the Cooper-pair momentum with SC1, SC2, and SC3. This SC5 state exhibits discontinuous phase boundaries with both SC4 and SC1, resulting in a triple point of SC1, SC4, and SC5. 
In contrast to the SC1-SC4 transition, the first-order phase transition line between SC1 and SC5 terminates at a critical point, analogous to the behavior observed in Fig.~\ref{fig:PD_Zeeman}. Instead, there exists the first-order phase transition line between SC4 and SC5 even in a large $\so$ region.

\section{Conclusion}
In this paper, we have investigated the possibility of the finite-momentum superconductivity in the 2D $d$-wave altermagnet with the RSOC, mainly solving the linearized gap equation. We have revealed two superconducting states in terms of whether the dominant pairing mechanism is the inter-band or intra-band pairing, depending on the strength of the altermagnetic spin-splitting $\AM$ and the RSOC $\so$. 
Both of the superconductivity favor the finite-momentum states in the case of the in-plane N\'{e}el vector of the altermagnet, while the finite-momentum state is possible only for the inter-band pairing in the case of the out-of-plane N\'{e}el vector.
The origin of the difference between the in-plane and the out-of-plane altermagnet is explained by the coupling/decoupling of the N\'{e}el vector and the RSOC.
Furthermore, in terms of a comparison with the uniform magnetic field, we have clarified two unique features of the superconducting states in the in-plane altermagnet with the RSOC.
One is the expansion of the inter-band pairing region in the phase diagram for large $\so$, and the other one is that the intra-band pairing can be stabilized by Cooper pairs on both the outer and inner FSs. These features can be understood through the lens of the nesting condition and the anisotropic deformation of the two FSs affected by the in-plane altermagnetic spin-splitting and the RSOC.
We have also performed the self-consistent calculation, and revealed the whole property of the phase diagram for the in-plane altermagnet.

Although we have only dealt with the spin-singlet superconducting order parameter, in general the spin-singlet and the spin-triplet superconductivity can be mixed in noncentrosymmetric systems.
It would be worthwhile to study parity-mixed finite-momentum superconducting states in altermagnets in the context of the topological superconductivity\cite{SCinAM_topoSC, TopoSC_Lee_Qian_Yang_2024, topoSC_Li_2024, BFS_in_AM_Hong_Park_Kim_2024,Li_Liu_2023_topoSCinAM}.
In addition, the efficiency and the sign of the SDE are also important issues for applications of altermagnets.
The sign change or the high efficiency of the SDE is expected when a phase transition of the superconductivity takes place~\cite{daido_diode, PhysRevB-daido-diode,PerfectSDE_Chakraborty_Black-Schaffer_2024}.
Therefore, it is one of the future issues to clarify the properties of the SDE in the vicinity of the first-order transitions discussed in this work.

\begin{acknowledgment}

%\acknowledgment
This work was supported by JSPS KAKENHI Grant Nos.~JP24K17000 and JP23K22492.

\end{acknowledgment}
\appendix

\section{Orientation of Cooper-pair momentum}
\label{sec:orientation_of_q}
\begin{figure*}[tb]
    \centering
    \includegraphics[width=\linewidth]{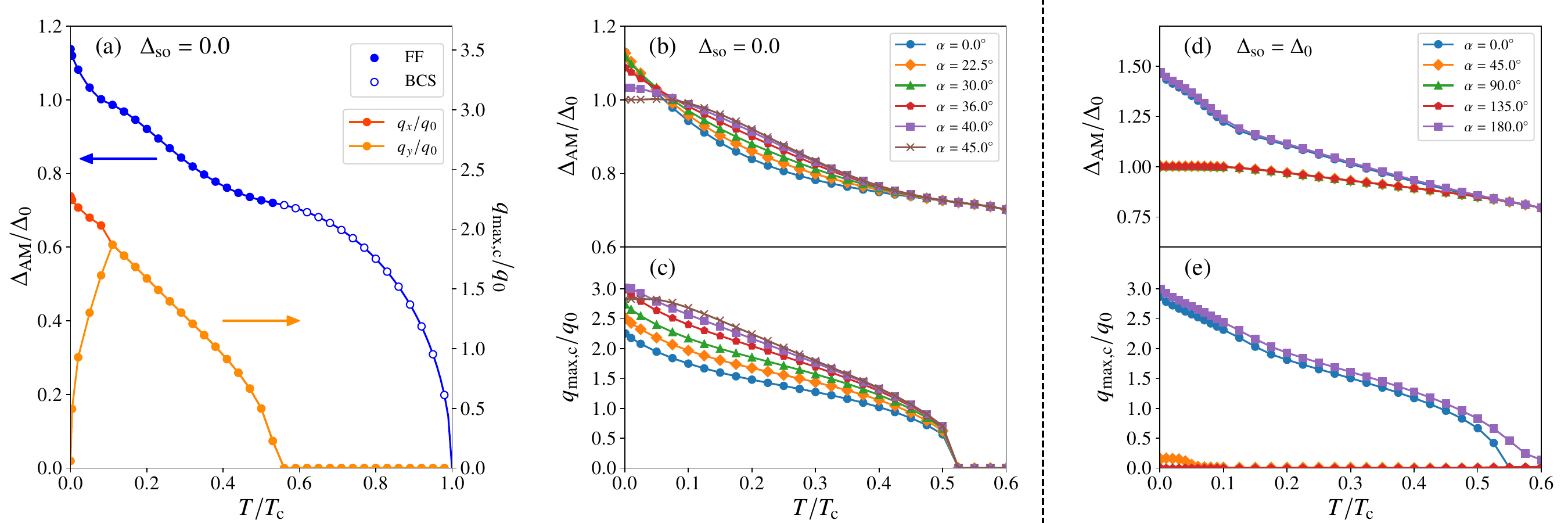}
    \caption{(a) Phase diagram in the $T$-$\AM$ plane and temperature dependence of $\bm{q}_{\mathrm{max,c}}$ without the RSOC. The blue solid (open) circles account for the superconducting state whose momentum is finite (zero). (b, d) Phase diagrams in the $T$-$\AM$ plane and (c, e) $T$-dependencies of $|\bm{q}_{\mathrm{max,c}}|$ for several fixed orientations for $\so/\Delta_0=0.0$ (b, c) and 1.0 (d, e). The orientation of $\bm{q}_{\mathrm{max,c}}$ is denoted by $\alpha = \tan^{-1}(q_y/q_x)$}
    \label{fig:degre_dep_matome}
\end{figure*}
In this Appendix, we present the results on the orientation of the Cooper-pair momentum. Firstly, we focus on the $d$-wave superconductors in the altermagnet without the RSOC. 
We show the $T$-$\AM$ phase diagram in Fig.~\ref{fig:degre_dep_matome}(a) and discuss the temperature dependence of $\bm{q}_{\max,\mathrm{c}}$. 
In Fig.~\ref{fig:degre_dep_matome}(a), the blue solid circles ($T \lesssim 0.55\Tc$) account for the finite-momentum superconductivity (FF-state), while the open circles do the conventional BCS pairing, along the phase boundary.
Focusing on the variation of $\bm{q}_{\mathrm{max,c}}$, one can see that the orientation of $\bm{q}_{\mathrm{max,c}}$ depends on the temperatures, and it gradually changes from $45^{\circ}$ to small angles with temperature decreasing.
To investigate details, we show phase diagrams 
for several values of the orientation of the Cooper-pair momentum $\alpha = \tan^{-1}(q_y/q_x)$
in Fig.~\ref{fig:degre_dep_matome}(b,c).
For temperatures above $T^*\approx0.1\Tc$, the momentum $\bm{q}_{\max}$ that realizes the largest $\AM$ is along $45^{\circ}$.
As temperature decreases, the most stable orientation continuously changes from $45^{\circ}$ to smaller angles. Such a change  results from the competition of the $T$-dependent $\Delta_{\mathrm{AM,c}}$ for different $\alpha$:
in Fig.~\ref{fig:degre_dep_matome}(b), when $\bm{q}$ is along the antinode direction of the $d$-wave superconductivity, $\Delta_{\mathrm{AM,c}}$ shows an upturn behavior with decreasing temperature, while when $\bm{q}$ is along the node direction, the saturated behavior is observed. 
These results at low temperatures are consistent with the previous works\cite{zero-field_Chakraborty_Black-Schaffer_2024,Soto-Garrido_Fradkin_2014} on the $d$-wave superconductors in the altermagnet. 
We additionally note that the similar temperature dependence of $\bm{q}_{\mathrm{max,c}}$ is reported in a 2D $d$-wave superconductor with a uniform magnetic field\cite{Vorontsov_Sauls_Graf_2005,Shimahara_1998}.

Figure~\ref{fig:inplane_T=0.01}(a) shows that only $q_x$ remains finite while $q_y$ is zero when both effects of the RSOC and the altermagnetic spin-splitting is finite.
To see how the RSOC affects the orientation of $\bm{q}_{\mathrm{opt}}$, we show the phase diagrams with $\so=\Delta_0$ for several values of $\alpha$ in Fig.~\ref{fig:degre_dep_matome}(d),
where the state with the momentum along the $k_x$-direction is confirmed to be the most stable in the whole temperature region where a FF-state is favored.
One can see from Fig.~\ref{fig:degre_dep_matome}(e) that $q_{\mathrm{max,c}}$ for $\alpha = 45^{\circ}$, $90^{\circ}$, and $135^{\circ}$ becomes zero or much smaller than those for $\alpha = 0^{\circ}$ and $180^{\circ}$, which means that the pairing states are similar to the BCS state.
Hence, the upturn behavior in the low temperature region cannot be observed for those angles [Fig.~\ref{fig:degre_dep_matome}(d)].
The slight difference of $\Delta_{\mathrm{AM,c}}$ and $q_{\mathrm{max,c}}$ between $\alpha = 0^{\circ}$ and $180^{\circ}$ represent that $\bm{q}$ and $-\bm{q}$ are no longer equivalent. 
The previous studies have shown that in a system with the RSOC and the in-plane magnetic field, the finite-momentum superconductivity whose momentum is perpendicular to the magnetic field gets stable\cite{Kaur_Agterberg_Sigrist_2005,Agterberg-helicalstripe-2007}. Therefore, in terms of the direction of the momentum, our result is consistent with the situation of the helical state.

\section{Results in a uniform magnetic field}
\label{sec:uniform magnetic field}
We here show the results in a uniform magnetic field case as a comparison to those in the altermagnet case in the main text. Instead of Eq.~\eqref{eq:hvector}, we consider the following momentum-dependent spin-splitting:
\begin{equation} \label{eq:hvector_B}
    \bm{h}(\bm{k}) = \so\hat{\bm{k}}\times\bm{e}_z + B\bm{n},
\end{equation}
where $\bm{n}$ and $B$ are the direction and the strength of the magnetic field, respectively. We set $\bm{n}=\bm{e}_y$ and consider the combined effect of the RSOC and the in-plane magnetic field. The energy dispersion for this case is given as
\begin{equation}
    E^{\mathrm{in}}_{\lambda}(\bm{k})=\xi_k+\lambda\sqrt{(-\so\hat{k}_x+B)^2+\so^2\hat{k}_y^2}.
\end{equation}
We consider the $s$-wave superconductor, and calculate physical quantities in the same framework of the self-consistent method as in Sec.~\ref{sec:self-consistent_AM}.

\begin{figure}[tb]
    \centering
    \includegraphics[width=\linewidth]{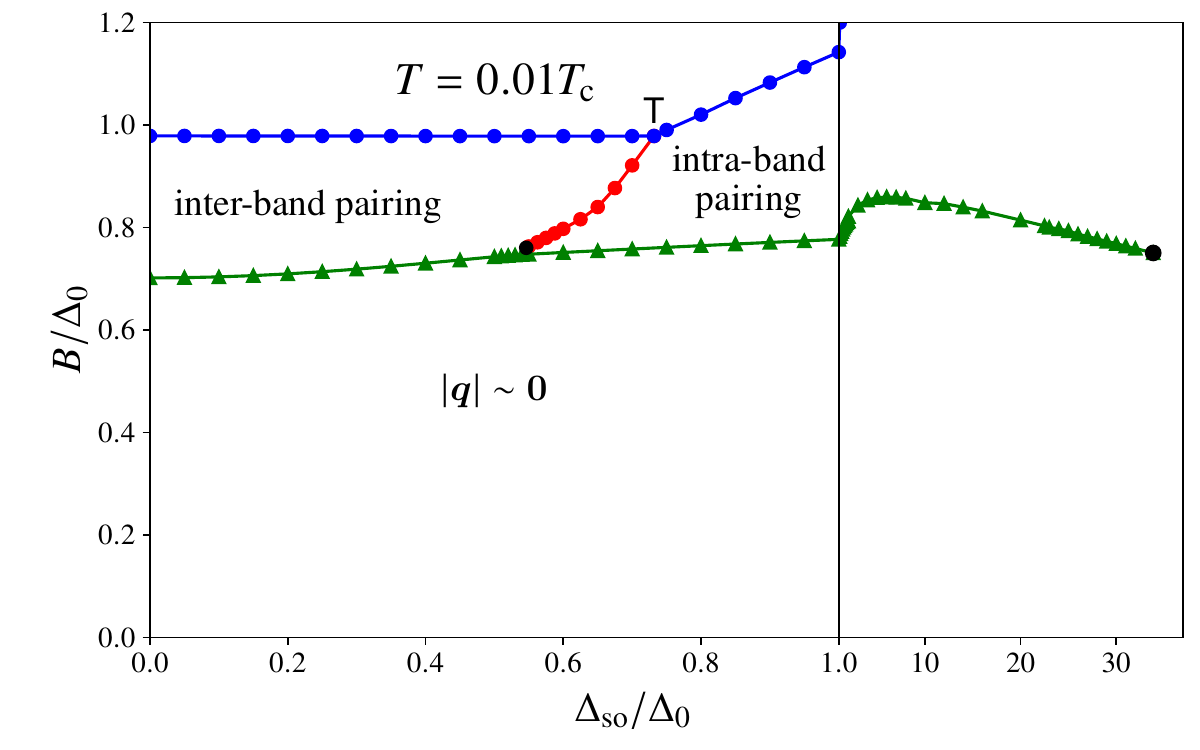}
    \caption{Phase diagram for the in-plane uniform magnetic field in the $\so$-$B$ plane at $T=0.01\Tc$. Definitions of the lines are the same as those in Fig.~\ref{fig:PD_BdG}(a). Note that there is a difference of the scale of $x$-axis between the region $0\le\so/\Delta_0\le1.0$ and region $1.0<\so/\Delta_0$. The black points denote the critical points between two adjacent phases.}
    \label{fig:PD_Zeeman}
\end{figure}

\begin{figure}[tb]
    \centering
    \includegraphics[width=\linewidth]{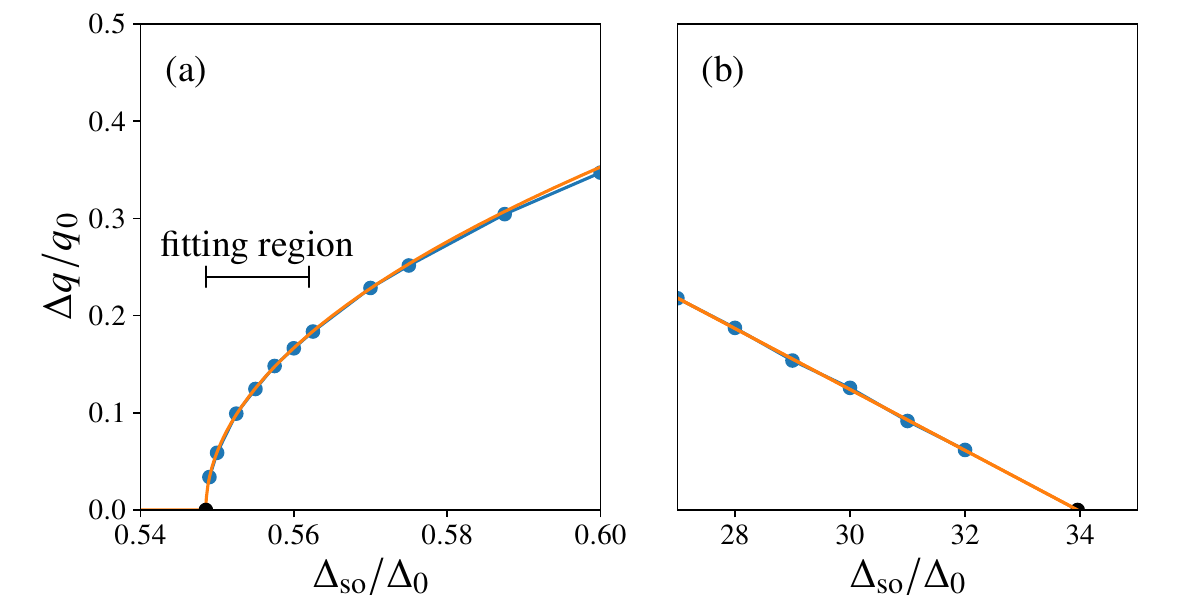}
    \caption{$\Delta \bm{q}$ as a function of $\so$ along the first-order phase transition lines. The left (right) panel corresponds to the boundary between the inter-band pairing state and the intra-band pairing state (the $|\bm{q}|\sim0$ state and the intra-band pairing state). The orange lines indicate the fitting curve.}
    \label{fig:Delta_q}
\end{figure}

We show the phase diagram for the in-plane uniform magnetic field in the $\so$-$B$ plane at $T=0.01\Tc$ in Fig.~\ref{fig:PD_Zeeman}. 
For large $\so$ ($\so>2\Delta_0$), we conduct the calculation in the $\bm{k}$ space satisfying $|\varepsilon_n(\bm{k},\bm{q})|<\omega_{\mathrm{D}}$.
The triple point $\mathsf{T}$ on the second transition line is located at $(\so/\Delta_0,\,B/\Delta_0)\approx(0.732,0.979)$. 
The phase diagram consists of three superconducting states: (i) the small $\bm{q}$ state labeled ``$|\bm{q}|\sim0$'',  (ii) the inter-band pairing state, and (iii) the intra-band pairing state.
We note that the first-order transition lines between the intra-band pairing state and the inter-band pairing state and between the $|\bm{q}|\sim0$ state and the intra-band pairing state terminate at the critical points  $(\so/\Delta_0,B/\Delta_0)\approx(0.55, 0.76)$ and $(33.9, 0.75)$, respectively.
These points are obtained as follows.
We introduce $\Delta \bm{q}=|\bm{q}_{\mathrm{min,1}}-\bm{q}_{\mathrm{min,2}}|$, where $\bm{q}_{\mathrm{min},1}$ and $\bm{q}_{\mathrm{min},2}$ are the momenta of two superconducting phases adjacent on a first-order transition line.
The critical point is determined so that $\Delta\bm{q}$ becomes zero, and we estimate the critical point by the curve fitting. 
Figure~\ref{fig:Delta_q} shows
$\Delta \bm{q}$ as a function of $\so$ and details of the curve fitting.
Note that the momentum of the $|\bm{q}|\sim0$ state is no longer small but comparable to that of the intra-band pairing near their critical point.
The phase diagram similar to that in Fig.~\ref{fig:PD_Zeeman} is also reported in the previous works\cite{topological-metal-liang-fu-2021,helical_Zwicknagl_Jahns_Fulde_2017}.

\begin{figure}[tb]
    \centering
    \includegraphics[width=\linewidth]{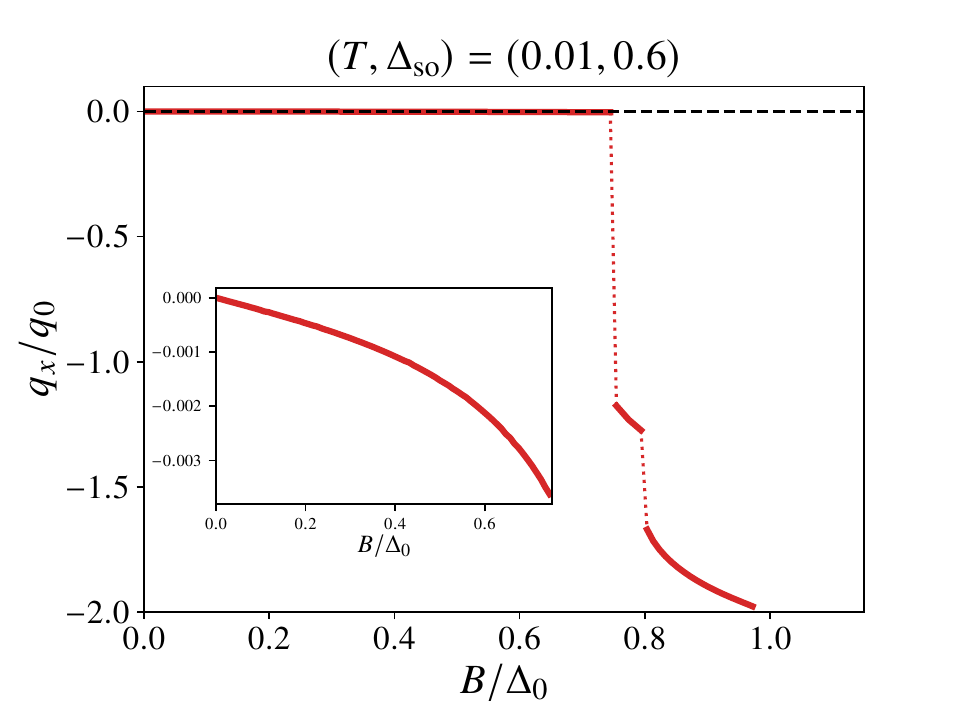}
    \caption{$B$ dependence of  $\bm{q}_{\mathrm{opt}}$ for $\so=0.6\Delta_0$. The inset shows an enlarged view in the region $0\le B/\Delta_0\le0.75$.}
    \label{fig:qopt_zeeman}
\end{figure}

In Fig.~\ref{fig:qopt_zeeman}, $\bm{q}_{\mathrm{opt}}$ is plotted as a function of $B$ at $\so=0.6\Delta_0$. Discontinuous changes of $\bm{q}_{\mathrm{opt}}$ are observed at $B\approx0.75\Delta_0$ and $B\approx0.8\Delta_0$. 
They correspond to the two first-order transitions.
The former one is from the $|\bm{q}|\sim0$ state to the intra-band pairing state, and the latter one is from the intra-band pairing state to the inter-band pairing state. 
Especially, $\bm{q}_{\mathrm{opt}}$ is always along the $-k_x$ direction in the whole region of the intra-band pairing state, in contrast to the case of the altermagnet.

\begin{figure}[tb]
    \centering
    \includegraphics[width=\linewidth]{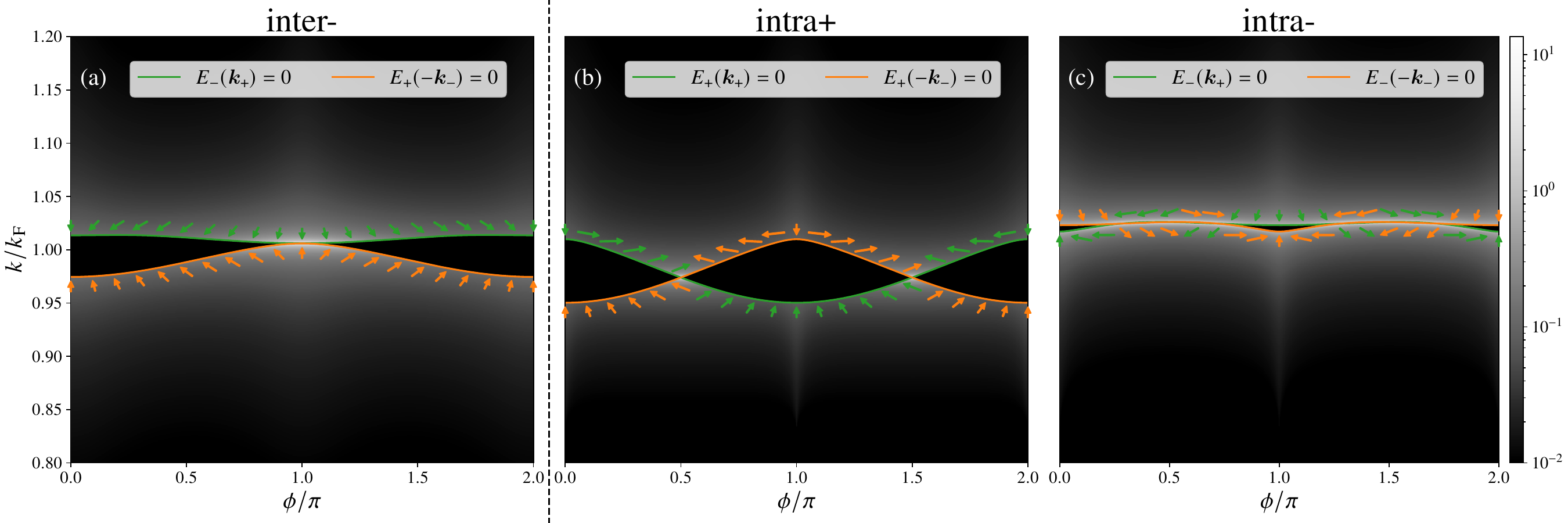}
    \caption{FSs for two parameter-sets on the second-order transition line in Fig.~\ref{fig:PD_Zeeman}. (a) The inter-band pairing for $(\so/\Delta_0,B/\Delta_0,q_x/q_0)=(0.6,0.98,-1.94)$ and (b, c) the intra-band pairing for $(\so/\Delta_0,B/\Delta_0,q_x/q_0)=(2.0,1.64,-2.68)$.}
    \label{fig:Zeeman_nesting}
\end{figure}

Here, we make a discussion on the nesting condition of FSs of the uniform magnetic field case as a comparison to the cases of the altermagnet, showing the FSs in Fig.~\ref{fig:Zeeman_nesting}.
We choose two sets of parameters $(\so/\Delta_0,B/\Delta_0,q_x/q_0)=(0.6,0.98,-1.94)$ for panel (a), and $(\so/\Delta_0,B/\Delta_0,q_x/q_0)=(2.0,1.64,-2.68)$ for panels (b) and (c).
Both of them correspond to the second-order transition point in Fig.~\ref{fig:PD_Zeeman}.
For the inter-band pairing [Fig.~\ref{fig:Zeeman_nesting}(a)], compared to the case of the altermagnet shown in Fig.~\ref{fig:nesting_FS}(b), the FSs are not sufficiently nested to stabilize the superconducting states in larger $B$. Hence, there is no enhancement of the critical field with $\so$ for the inter-band pairing, while the enhancement results in the reentrant region in the case of the altermagnet.
For the intra-band pairing [Figs.~\ref{fig:Zeeman_nesting}(b) and \ref{fig:Zeeman_nesting}(c)], it can be found that only the ``intra$-$'' pairing, Cooper pairs on the outer FS, has a better nesting condition than the ``intra$+$''. 
This corresponds to the mechanism of the intra-band pairing that the momentum is optimized so that the nesting condition becomes good only in the FSs with larger density of states.

\bibliographystyle{jpsj}
\bibliography{reference}
\end{document}